\documentclass[11pt]{article}

\usepackage{amsmath}
\usepackage{amsthm}
\usepackage{amssymb}
\usepackage{amscd}
\usepackage{amsfonts}
\usepackage{amsbsy}
\usepackage{setspace}
\usepackage{graphicx}
\usepackage{calc}
\usepackage{verbatim}
\usepackage{algorithmicx}
\usepackage{verbatim}
\usepackage{euscript}
\usepackage{color}
\usepackage{titletoc}
\usepackage{changepage}
\usepackage{framed}
\usepackage{bm}
\usepackage{authblk}
\usepackage{graphicx,subfigure}
\usepackage{multirow}
\usepackage{enumitem}
\usepackage{tikz}
\usepackage{physics}
\usepackage{changepage}
\usepackage{parskip}
\usepackage[T1]{fontenc}
\usepackage[protrusion=true,expansion=true]{microtype} 
\usepackage{color,soul}
\usepackage{tabu}

\definecolor{light-gray}{gray}{0.9}
\newtheorem {example} {Example}

\newcommand{\p}{\partial}
\newcommand{\R}{\ensuremath{\mathbb{R}}}

\newcommand{\PRC}{{\rm PRC}}
\newcommand{\SRC}{{\rm SRC}}
\newcommand{\PCC}{{\rm PCC}}
\newcommand{\CC}{{\rm CC}}
\newcommand{\EE}{{\rm EE}}
\newcommand{\PRCC}{{\rm PRCC}}
\newcommand{\SRCC}{{\rm SRCC}}
\newcommand{\SI}{{\rm SI}}
\newcommand{\I}{{\rm I}}
\newcommand{\E}{{\rm E}}
\newcommand{\V}{{\rm V}}
\title{\bf \huge Sensitivity analysis methods\\ in the biomedical sciences}
\author{\bf George Qian and Adam Mahdi }
\affil{Institute of Biomedical Engineering \\
University of Oxford}

 \date{}

\begin{document}
%%%%%%%%%%%%%%%%%%%%%%%%%%%%%%%%%%%%%%%%%

\maketitle

\begin{abstract}
\noindent
Sensitivity analysis is an important part of a mathematical modeller's toolbox for model analysis. In this review paper, we describe the most frequently used sensitivity techniques, discussing their advantages and limitations, before applying each method to a simple model. Also included is a summary of current software packages, as well as a modeller's guide for carrying out sensitivity analyses.  Finally, we apply the popular Morris and Sobol methods to two models with biomedical applications, with the intention of providing a deeper understanding behind both the principles of these methods and the presentation of their results.
\end{abstract}


\tableofcontents

%%%%%%%%%%%%%%%%%%%%%%%%%%%%%%%%%%%%%%%%%
\section{Introduction}

Sensitivity Analysis (SA) can be defined as the study of how uncertainty in a model's output can be apportioned to different sources of uncertainty in the model input \cite{Saltelli2008}. Note that SA differs from uncertainty analysis (UA) which, instead, characterises the uncertainty in the model output in terms of, for example, the empirical probability densitiy or confidence bounds \cite{Saltelli2000, Saltelli2019b}. In other words, UA asks how uncertain the model  output is, whereas SA aims to identify the main sources of this uncertainty \cite{Saltelli2010a}.  The analysis of a mathematical model can greatly benefit from including SA. Some of the common applications of SA include model reduction, inference about various aspects of the studied phenomenon or experimental design.

In the biomedical sciences and biology, SA is especially important for several reasons. Biological processes are inherently stochastic \cite{Lipniacki2006} and the collected data are subject to uncertainty \cite{White2004, Blower1994, Geris2016}. Also, while mathematical models  are important tools for formulating and testing hypotheses about complex biological systems \cite{Kitano2002b, Voit2006, Mahdi2013b}, a major obstacle confronting such models is that they typically have a large number of free parameters whose values can affect model behaviour and its interpretation. It has been observed that although high-throughput methods are well-suited for discovering interactions, they remain of limited use for the measurement of biological and biochemical parameters \cite{Maerkl2007, Gutenkunst2007}. Model parameters can also be approximated collectively through data fitting, rather than direct measurement \cite{Lillacci2010}. However, this often leads to large parameter uncertainties if the model is unidentifiable. SA methods can be used to ensure identifiability, a property which the model must satisfy for accurate and meaningful (unique) parameter inference, given the measurement data.

There have been many studies of SA techniques and their implementation. It is worthwhile briefly mentioning some reviews. These tend to include  applications to some specific area of research, such as complex kinetic systems \cite{Turanyi1990}, environmental models  \cite{Hamby1994, Norton2008, Pianosi2016}, building energy analysis \cite{Wei2013}, radioactive waste \cite{Helton1993}, hydrogeology \cite{Wainwright2014a}, operations research \cite{Borgonovo2015}, reliability analysis \cite{Aven2010} and system biology \cite{Zi2008, Marino2008a}.  In addition, there are more general reviews \cite{Saltelli2006} and several textbooks introducing the field, which tend to focus on global methods \cite{Eslami2013, Saltelli2000, Saltelli2002, Saltelli2008}.

While reviews and books on SA do already exist, here we provide an elementary introduction to sensitivity methods, together with some practical examples with a biomedical focus. Also included is an overview of each method, the settings where it is advantageous to apply these methods, as well as where their limitations lie. We then apply each method to a simple example problem, illustrating the results. There follows, for the benefit of readers who wish to conduct their own SA, a summary of computational software implementing different techniques, as well as a practical workflow. Finally, we apply the Morris and Sobol methods, two popular techniques, to both an algebraic and a time-dependent biomedical model. Our aim is to introduce the readers to SA techniques, showing how to choose the most suitable approach for the problem at hand, as well as appropriate practices for SA implementation.

%%%%%%%%%%%%%%%%%%%%%%%%%%%%%%%%%%%%%%%%%
\section{Basic definitions and concepts}
This section introduces some basic definitions and concepts used in the context of SA. We will refer to the terminology introduced here in later sections.

%---------------------------------------------------
\subsection{Input factors and outputs}
When we refer to a {\it mathematical model}, we understand a relationship of the form
\begin{equation}
\label{ModelForm}
Y=f(X_1,X_2,...X_n),
\end{equation}
where $Y$ is the {\it model output} and $X_1,\ldots,X_n$ the {\it model inputs}. The output can be either single-valued or a vector. This review primarily focuses on models of this form. To be consistent with the terminology in the literature, we will also refer to the $i$th input, $X_i$, as the {\it i}th model {\it factor}.

 \begin{figure}[t!]
\centering
      \includegraphics[width=0.9\linewidth]{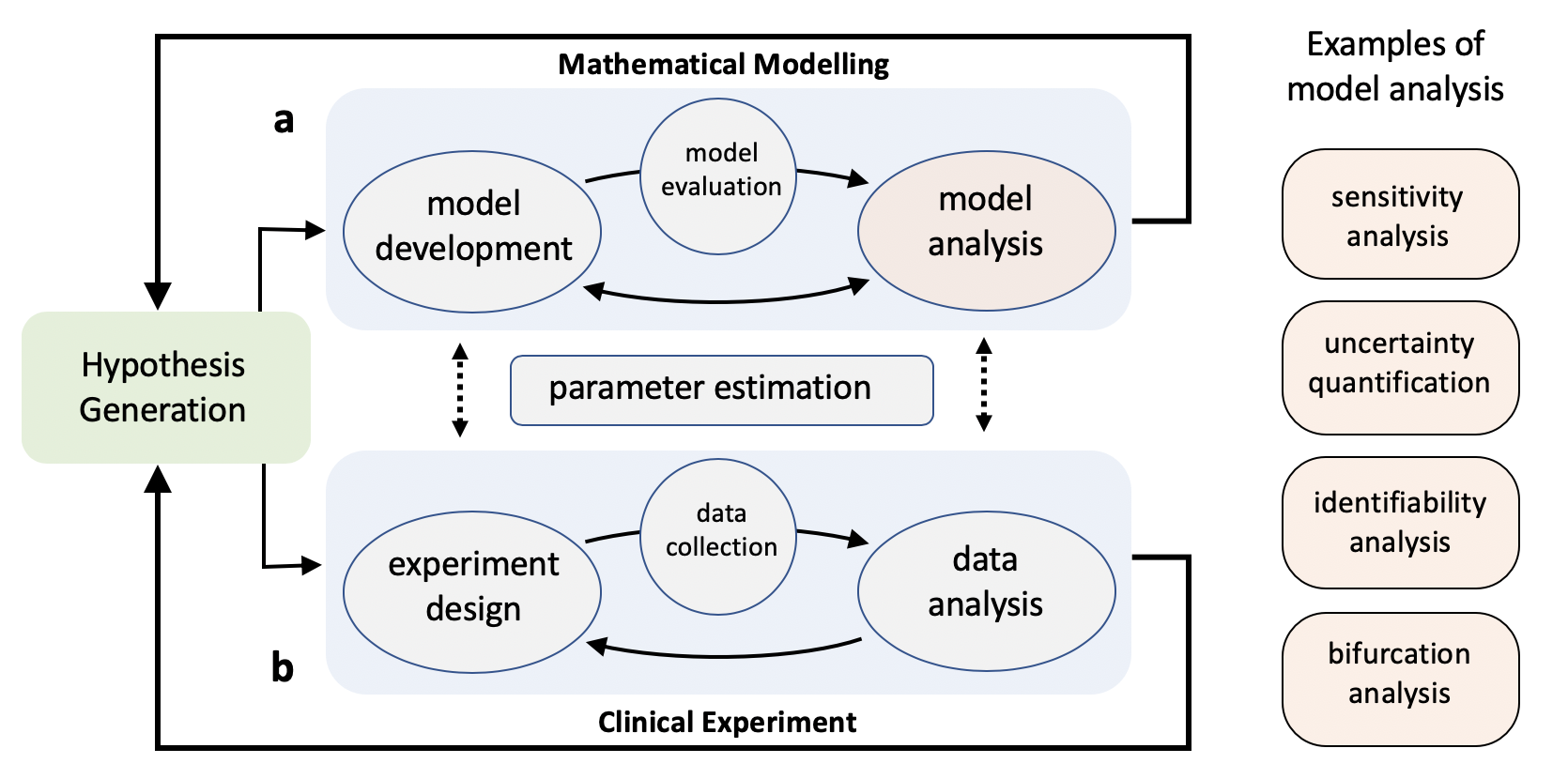}
  \caption{{\bf Typical workflow for experimentation and modelling in biomedical sciences.}  Sensitivity analysis is an integral part of the mathematical modelling cycle and is placed within model analysis module. The flowchart (adapted from \cite{Tanaka2013}) shows how modelling and experiments can be conducted to investigate a phenomenon of interest.  After making a hypothesis, we conduct experiments and implement mathematical models. Model analysis follows the latter and this may include structural and practical identifiability, uncertainty quantification and sensitivity analyses, which is focus of the this review.}
  \label{FlowChart}
\end{figure}

%-----------------------------------------------------
\subsection{Experimentation and modelling in biomedical sciences}

Mathematical modelling of biological or physiological systems typically begins by setting the scope of the study and defining the research question, or hypothesis. This includes identification of the model structure, such as the inputs and outputs of the system and any interactions between various components, a process informed by the underlying biomedical phenomenon \cite{Keener2009}. We note that in this structure the outputs encode information about the normal and affected states. The choice of the most appropriate mathematical modelling approach, which may range from a simple algebraic equation to a complex time-dependent system, is dependent upon the nature of the biological problem under investigation. 

The parallel between the workflow for clinical experiments and mathematical modelling is illustrated in Figure \ref{FlowChart}. The  mathematical modelling workflow (Figure  \ref{FlowChart}a) allows us to develop the model and subsequently simulate conditions which may not be feasible in a real-world setting. The analogous clinical experiments (Figure  \ref{FlowChart}b) yield data which are subsequently analysed and the results fed back to either modify the experiment or draw conclusions about the research question.  Clinical data may also be used, if available,  to further calibrate the model through parameter estimation. 

After completing the initial mathematical model, model analysis is performed \cite{Kitano2002}. This includes a variety of techniques such as practical identifiability \cite{Miao2011, Olufsen2013, Saccomani2016}, stability \cite{Mader2015} and bifurcation analysis \cite{Mahdi2017b}, as well as SA, which is the main topic of this paper. Within the mathematical modelling workflow, it is possible to perform model analysis even without evaluating the model by using, for instance, structural identifiability techniques \cite{Villaverde2016, Mahdi2014}.

 %----------------------------------------------------- 
 \subsection{Aims of SA}
 Before proceeding, it is important to understand what SA can achieve. In the literature, we commonly identify the following aims:
\begin{itemize}
\item {\it Ranking} involves ordering the inputs in terms of their effect on the variability of model output. Those ranked more highly then become the focus of experimental or numerical estimation, since the elucidation of these input values results in the greatest reduction of the output uncertainty.
\item {\it Screening} establishes which model inputs are unidentifiable, i.e. those which have little or no effect on the variability of mode output. This allows us to reduce the model by setting these inputs to fixed values.
\item {\it Mapping} observes the effect of the inputs on the output. This may involve determining the parameter values for which the model is stable, reaches its maximum output, or reaches some optimal set of values, for instance. 
\end{itemize}

There is no method suitable for all three SA applications and so careful selection of the method must occur, depending on the aim. To do this requires classification of the various SA methods, and so necessitates the introduction of some terminology. Since most sensitivity analysis techniques can be used for ranking or screening but not mapping, the review focuses on the first two applications.

%-----------------------------------------------------
\subsection{Local and global SA}
SA methods can also be classified by determining whether they explore the input space at one point or a larger region of it. 
\begin{itemize}
\item {\it Local} methods examine the sensitivity of the model inputs at one specific point in the input space. 
\item {\it Global} methods, on the other hand, take the sensitivities at multiple points in the input space, before taking some measure of the average of these sensitivities. This averaged value then represents the influence that the input exerts on the output uncertainty. 
\end{itemize}

While the simplicity and low computational cost of the local methods ensure their popularity, they tend to be informative only when the model is linear, as the sensitivity at any point can then be extrapolated to those further away. Extrapolation of a nonlinear model, however, may  lead to incorrect conclusions \cite{Saltelli2008}. Many global methods can also give some indication of each input's interactions with other input(s), furthering the case for their implementation on nonlinear models.

%-----------------------------------------------------
\subsection{One-at-a-time and Many-at-a-time SA} \label{sec:OATMAT}
Each sensitivity technique requires the evaluation of the model using different sets of input values. Hence, another way to classify these techniques is by simply counting how many input values change for each successive model simulation.
\begin{itemize}
\item {\it One-at-a-time} (OAT) methods change only one input per model evaluation. 
\item {\it Many-at-a-time} (MAT) methods operate by changing two or more inputs simultaneously, which we refer to as {\it many-at-a-time} (MAT) methods.
\end{itemize}

Most OAT techniques start from a baseline set of input values at which the model is known to converge. Hence, changing the value of one input reduces the chance of the model evaluation failing due to, for instance, instabilities or numerical error. Aside from their resilience to convergence issues, another advantage of OAT methods is that, if the model evaluation of a set of baseline values differs from that when one of the inputs is changed, then we can hold responsible that particular input for causing such a difference: it must have some effect on the output, at least at that location in input space. OAT methods, however, suffer from some serious limitations, notably their difficulties in analysing nonlinear models (see Section \ref{OATLimitationsSection}).  

Global sensitivity techniques can be either OAT or MAT, while local methods can only be OAT-based. Although MAT are more likely to evaluate the model at input values where it is unstable, and are often computationally more expensive, they are preferred for analysis of nonlinear functions as they explore a larger proportion of factor space \cite{Saltelli2000, Saltelli2010a}.

%-----------------------------------------------------
\subsection{Sampling strategies} \label{samplingsection}
\medskip
Frequently, sensitivity indices cannot be calculated analytically. Instead, numerical approximations are computed, necessitating the generation of input samples through a process called  sampling. This section briefly examines some common sampling procedures. These are required for the {sampling-based methods} (Section \ref{corregress}), which rely on the relationship between the sampled inputs and their corresponding outputs \cite{McKay1979, Iman1981}. The procedures can also be used for the sampling stage of other sensitivity techniques, including the Sobol method (Section \ref{sobol}).

\begin{itemize}
\item In {\it Random sampling} a number taken from the interval $[0, 1]$ is associated with each of $n$ model factors to form a vector \cite{Fishman1996, Barry1996}.  In practice, this is done using pseudorandom number generators (by reproducible algorithmic processes). 
The process is repeated to generate $k$ vectors of dimension $n$. Random sampling does not record the history of previous points and therefore it is possible for points to accumulate in some region \cite{Saltelli2000}. The issue with taking clustered samples is that the model approximations become less reliable in regions further away from this cluster, due to nonlinearities, for instance. Hence, it is good practice to use sampling strategies that fill the factor space more evenly, which we discuss next \cite{Forrester2008}.

\item  {\it Latin hypercube sampling} generates near-random samples  from a multidimensional distribution, ensuring that samples have low discrepancy, i.e.  are more evenly distributed across the factor space \cite{Helton2003}.  It is a generalisation (to any number of  dimensions) of the concept of a Latin square, which is a grid containing sample positions in such a way that there is only one sample in each row and column. When implementing the Latin hypercube method, the samples are generated sequentially and one must record, in contrast to the random sampling method, in which row and column the previous samples have been placed \cite{Seaholm1988}.

\begin{table}
%\begin{adjustwidth}{-2.7cm}{}
\centering

\small

%\footnotesize
\begin{tabu}{llllllll}
\tabucline[0.7pt] {-}	
Method                 	&Section				& Type	    	& Inter- 	         				& Non-			& Cost			&Evalu- 				&Aim\\
		                 &		 			& 		   	& actions 	         				& linearities 		&  	                 	&ations                  	     	&  \\ 
\hline
One-way                	&\ref{Sec:oneway}	        	& Local           	& No           					& No            	 	 &Low 	         	&OAT        			&Rank$^{**}$   \\
Multi-way              	& \ref{Sec:multiway}		& Local           	& No          					& No            		 & Low		     	&MAT				&Rank$^{**}$ 	\\
Local-derivative      	&\ref{Sec:local}			& Local           	& No           					& No             		 & Low    			&OAT				&Rank$^{**}$	\\
Morris                 	&\ref{Sec:morris} 		& Global          	& Yes          					& Yes 	                 & Medium  		&OAT       				&Screen \\
Sobol                  	&\ref{sobol}		     	& Global         	&Yes          					&Yes                          & High  		 	&MAT     				&Rank, Screen \\
FAST/eFAST            	&\ref{Sec:FAST}		& Global          	&Yes          					& Yes                         & High   			&MAT    				 &Rank, Screen\\
DGSM                      	&\ref{Sec:DGSM}		& Global           &Yes          					&Yes                          & Medium  		&OAT       				&Screen\\
Sensitivity index        &\ref{Sec:SI}	        		& Local             & No           					& No             		 & Low  			&OAT       				&Rank$^{**}$\\
Importance index     &\ref{Sec:Importance}	& Global           & No          					& Yes              		 & Low 		       	&OAT        			&Rank\\
$\CC_i$                    &\ref{corregress}		& Global          	& No           					& No             		 & Medium               	&OAT          			&Rank\\
$\SRC_i$                  &\ref{corregress}		& Global        	&No$^*$          					& No      			& Medium               	&OAT          			&Rank\\
$\PRC_i$                   &\ref{corregress}	      	& Global           	&Yes         					& No             		 & Medium               	&OAT           			&Rank\\
$\SRCC_i$                &\ref{corregress}		& Global        	& No          					& Yes                         &Medium                	&OAT           			&Rank\\
$\PRCC_i$                &\ref{corregress}		& Global        	& Yes         					& Yes                         &Medium                	&OAT           			&Rank\\
\tabucline[0.7pt] {-}
\end{tabu}
\caption{Summary of SA methods covered in this paper, stating whether they are local or global methods, their ability to handle nonlinearities in the model, detect interactions between factors, the computational cost and aim(s). 
$^*$The SRC can be modified to detect interactions between factors \cite{Robert2010}. 
$^{**}$ Note that using any local method for ranking is typically valid only in a small neighbourhood around the nominal values \cite{Saltelli2008,Saltelli2019b}.}
\label{MethodsSummary}
\end{table}

\item {\it Sobol sequence} is another quasi-random, low discrepancy sequence, typically resulting in a faster convergence and more stable estimates \cite{Sobol1967}. Quasi-random sequences ensure that the samples fill the space more uniformly than uncorrelated random samples \cite{Lemieux2009}. As a consequence of this, quasi-random sequences require fewer model evaluations for convergence \cite{Niederreiter1978}. Sobol sequences are especially useful when numbers are computed on a grid and it is unknown {\it a priori}  how fine the grid must be to obtain sufficiently accurate results \cite{Trandafir}.
\end{itemize}

%%%%%%%%%%%%%%%%%%%%%%%%%%%%%%%%%%%%%%%%%
\section{Review of SA methods}
With so many SA methods available, it is worth examining the most commonly applied methods individually. Our discussion of each method will conclude by its application to the following simple model
\begin{equation}
\label{ToyModel}
\ Y=X_1^2+X_2X_3+X_4,
\end{equation}
where $X_i$ are uniformly distributed in the interval $[0,1]$ model inputs (factors), that is,  $X_i\sim{\it U}[0,1], i=1,\ldots,4$ and $Y$ is the model output.

%-------------------------------
\subsection{One-way sensitivities}\label{Sec:oneway}
Perhaps the simplest and most commonly-used form of SA is the {\it univariate} or {\it one-way} method. This has been used to assess cost-effectiveness, with examples found frequently in the area of healthcare \cite{Wen2016,Wang2003a}.  The method works by changing one factor by a fraction of its nominal value, holding all other factors constant, and observing the resultant fractional change in the output. Since it is an OAT method, the one-way method is most suitable when the model is linear. The results can be displayed in a table or using a {\it tornado plot} (see Section \ref{GraphicalSection}), a type of bar chart comparing one-way sensitivity indices of a model's factors (see  Section \ref{GraphicalSection}) \cite{Eschenbach1992,Borgonovo2015}. 

\medskip

\begin{example}\upshape
Table \ref{OneWayTable} shows the results of a one-way SA on model  \eqref{ToyModel}. All factors are set to a value of $0.1$, before each is incremented by 10\%, with all other factors held constant. We then calculate the percentage increase in the output after each increment. In this particular case, changing factor $X_4$ causes the greatest output variation.
\end{example}

\begin{table}
\centering
\begin{tabu}{lcc}
\tabucline[1pt] {-}
\hline
Factor & 10\% increase & 10\% decrease \\
\hline

$X_1$  & 1.75\%                                   & -1.58\%                              \\
$X_2$  & 0.83\%                                  & -0.83\%                               \\
$X_3$  & 0.83\%                                  & -0.83\%                               \\
$X_4$  & 8.33\%                                   & -8.33\%                              \\
\hline
\tabucline[1pt] {-}
\end{tabu}
\caption{This table shows the results of a one-way SA conducted on model  \eqref{ToyModel}. We tabulate the percentage change in the output after increasing and then decreasing each factor by 10\%. Each factors is initially set to a value of 0.1.}\label{OneWayTable}
\end{table}

%---------------------------------------------------
\subsection{Multi-way sensitivities}  \label{Sec:multiway}
If, instead of varying one model factor, we vary two or more simultaneously and observe the change in model output, we would then be conducting a {\it multi-way} SA. A two-way SA, for example, examines the effect of changing two factors of interest. However, it remains local in nature and so is best used to analyse linear models.  
Multi-way techniques have been used to investigate the cost-effectiveness of healthcare routines, such as the screening of smokers for suspected lung cancer \cite{Mahadevia2003}, treatments for ischaemic stroke \cite{Fagan1998} and a potential vaccine for the human papillomavirus \cite{Sanders2003}.

\begin{table}[]
\centering
{\small
\begin{tabu}{lllll}
\tabucline[0.7pt] {-}
Factor & $X_1$         & $X_2$          	& $X_3$         	 & $X_4$           	\\
\hline
$X_1$  & N/A            & 2.52		& 2.52		& 10.08			\\
$X_2$  & 2.52 		& N/A            	&1.75		& 9.17			\\
$X_3$  & 2.52		& 1.75    		& N/A            	& 9.17			\\
$X_4$  & 10.08		& 9.17		&9.17		& N/A        		\\  \tabucline[0.7pt] {-}
\end{tabu}
\hspace{4ex}
\begin{tabu}{lllll}
\tabucline[0.7pt] {-}
Factor & $X_1$         & $X_2$          	& $X_3$         	 & $X_4$           	\\
\hline
$X_1$  & N/A            &  -2.42 		& -2.42 		&  -9.92 			\\
$X_2$  & -2.42		& N/A        	&-1.58		& -9.17  			\\
$X_3$  & -2.42  	& -1.58      	& N/A            	& -9.17 			\\
$X_4$  &  -9.92 	& -9.17		&-9.17 		& N/A        		\\  \tabucline[0.7pt] {-}
\end{tabu}
}
\caption{The results of a two-way SA conducted on model \eqref{ToyModel}. We tabulate the percentage change in the output after increasing and then decreasing two factors by 10\%. Each factor is initially set to a value of 0.1. Here, N/A stands for {\it not applicable}.}\label{MultiwayTable}
\end{table}

\medskip
\begin{example}\upshape
Table \ref{MultiwayTable} shows the results of applying a two-way SA to model \eqref{ToyModel}. First we set all factors to a baseline value; here, all factors are set to 0.1. We then apply a 10\% increase to two factors simultaneously and calculate the corresponding percentage change in the output. Analogous computations are performed for a 10\% decrease. In both cases, changing the pair $[X_1, X_4]$ creates the greatest magnitude of change in the output.
\end{example}

%-------------------------------
\subsection{Derivative-based local sensitivities} \label{Sec:local}
Traditionally, SA involved calculating the partial derivative of the output with respect to the input factor of interest. Mathematically, we can denote this as
\begin{equation}
\label{LocalSensEquation}
\ S_i^{method}=\gamma_i \frac{\partial Y}{\partial X_i}\Big\rvert_{X=x^*} 
\end{equation}
where $S_i^{method}$ is the sensitivity index of the $i$th factor, $x^*$ the point at which the derivative is evaluated, and $\gamma_i$ takes one of the following three forms, depending on the method used:

\smallskip

\noindent
{\it Absolute sensitivity index $S_i^{\rm abs}$} ($\gamma_i=1$). This gives us the rate of change of the output as $X_i$, alone, is altered. The index is best used when comparing factors with the same units and similar nominal values and variances. If a factor differs from the others in any of these three categories, the two indices below tend to be used instead.  
 
 \smallskip

\noindent
{\it Relative sensitivity index $S_i^{\rm rel}$} ($\gamma_i=X_i^0 / Y^0$).  Here $X_i^0$ is some nominal or reference value for the $i$th factor, and $Y^0$  some reference value for the output. The relative sensitivity index is a normalised measure and this enables comparisons between factors with different units or values at different orders of magnitude. However, one persisting limitation is that the spread, or standard deviation, of the respective factors remains unaccounted for. 

\smallskip

\noindent
{\it Variance sensitivity index $S_i^{\rm var}$} ($\gamma_i=\sigma_{X_i}/\sigma_Y$). Here  $\sigma_{X_i}$ and $\sigma_{Y}$ are the standard deviations of the factor $X_i$  and the output $Y$, respectively. Since the variance sensitivity index  $S_i^{\rm var}$ requires information about the spread of all the factors, it has been described as a hybrid local-global method \cite{Saltelli2002}.

\medskip

\begin{example}\upshape
Table \ref{TableLocal} gives the results of applying three local derivative-based sensitivity methods $S_i^{abs}, S_i^{rel}$ and $S_i^{var}$ to model \eqref{ToyModel}. The sensitivities are evaluated when all factors are set to 0.1. Note that the individual sensitivities depend on the point in factor space at which the methods are evaluated (with the exception of the linear additive term $X_4$).
\end{example}

\begin{table}[]
\begin{center}
\begin{tabu}{lccc}
\tabucline[.8pt] {-}
\hline
Factor & 	$S_i^{\rm abs}$									&	$S_i^{\rm rel}$									&$S_i^{\rm var}$\\
\hline
$X_1$  & 0.2                                                                                          	& 0.167			                                                                      	& 0.014                                                                \\
$X_2$  & 0.1                                                                                            	& 0.083                                                                                          	& 0.007                                                                                      \\
$X_3$  & 0.1                                                                                            	& 0.083                                                                                          	& 0.007                                                                                      \\
$X_4$  & 1.0                                                                                              	& 0.833                                                                        			& 0.071                                                                 \\
\hline
\tabucline[.8pt] {-}
\end{tabu}
\caption{ The results of local derivative-based sensitivities applied to model \eqref{ToyModel} calculated at 0.1. For the relative sensitivity index, $S_i^{\rm rel}$, we have $X_i^0=0.1$ and $Y_i^0=0.12$; and for the variance sensitivity index,  $S_i^{\rm var}$, $\sigma_{X_i}^2=0.083$ and $\sigma_Y^2=1.17$.}\label{TableLocal}
\end{center}
\end{table}

%---------------------------------------------------
\subsection{Morris method}\label{Sec:morris}
The Morris method \cite{Morris1991} can be viewed as an extension of the local derivative-based sensitivity measures of the previous section. This extension turns the Morris method into a global technique, making it one of the more widely applied sensitivity methods. 

Instead of taking the partial derivative of the output with respect to the factor of interest, say $X_i$, the Morris method approximates this derivative using a finite difference scheme. The resultant value is called the {\it elementary effect}, ${\rm EE}_i$ of the {\it i}th factor:
\begin{equation}
\label{MorrisMethodEq}
\EE_i=\frac{f(X_1,X_2,\dots,X_i+\Delta,\ldots,X_n )-f(X_1,X_2,\ldots, X_i, \ldots, X_n )}{\Delta}, 	
\end{equation}
where the increment or step-size, $\Delta$, is typically chosen to be ${n}/{(2(n-1))}$, with $n$ being the number of factors \cite{Saltelli2000}. 

Each of these factors is, first, rescaled to be uniformly distributed in the interval $[0,1]$. Starting from an initial base value, selected at random from this uniform distribution, one random factor is incremented or decremented and its elementary effect calculated. From this next value, another random factor is again incremented, and its elementary effect calculated and so on until we have calculated one elementary effect for each factor. The aim of the Morris method is to take the average of a number of elementary effects, each calculated at different points in factor space. Denoting this number by $r$, we would then require a total of $r$ elementary effects per factor. Hence, we repeat the process described above $(r-1)$ times to generate the remaining elementary effects. Each repetition, called a {\it run}, generates a set, or {\it trajectory}, of $n$ elementary effects (one per factor). Interested readers can find an algorithm for this process in \cite{Morris1991,Saltelli2002}. 

Having computed these $r$ elementary effects per factor, we find the average of their absolute values, $\mu_i^*$, and standard deviation of the signed values, $\sigma_i$:
\begin{eqnarray}
&&\mu^*_i=\frac{\sum_{k=1}^{r} \mid{\EE_i^k}\mid}{r} 	\label{MorrisMethodMuStar} \\ 
&&\sigma_i = \Big[\frac{\sum_{k=1}^{r} (\EE_i^k-\mu_i)^2 }{r-1} \Big]^{1/2} 			\label{MorrisMethodSigma} 
\end{eqnarray}
where $\EE_i^k$ denotes the elementary effect of the $i$th factor during the $k$th model evaluation and $\mu_i=\sum_{k=1}^{r} {\EE^k_i}/r$ is the mean of these elementary effects. 

The greater the value of $\mu_i^*$, the more the {\it i}th factor affects the model output, while the greater the $\sigma_i$ value, the more the factor is nonlinear or involved with interactions with other factors; a low $\sigma_i$, by contrast, indicates a linear, additive factor. While Morris' original paper \cite{Morris1991} used only $\mu_i$, Campolongo et al. \cite{Campolongo2007}  introduced the $\mu^*$ term. They observed that, by using their absolute values, elementary effects of different signs would not cancel each other out in \eqref{MorrisMethodMuStar}. However, the signed elementary effects are still used to calculate $\sigma_i$ in \eqref{MorrisMethodSigma}. The Morris method is computationally efficient and can be extended to deal with groups of factors \cite{Saltelli2002}. 

The sampling scheme used may affect the performance of the Morris method. For instance, the trajectory-based scheme can be modified by selecting only the trajectories with the largest geometric distance between their respective points \cite{Campolongo2007}. This is known as the {\it optimised trajectories} scheme. A related scheme is the {\it winding stairs} method, \cite{Jansen1994}. The difference between the winding stairs and trajectory-based schemes is that, while every point sampled by the former is linked to the same trajectory, in the latter, there are $r$ different sets of trajectories, each generated from a new initial base value. A different scheme that was first used in variance-based methods, known as the {\it radial design} \cite{Saltelli2002}, involves obtaining a trajectory where each point is only one step distant from the base point, which is a randomly-selected point in factor space. 
This was found to give better overall performance on a number of test functions than the trajectory-based scheme \cite{Campolongo2011}.   

The Morris method provides only semi-quantitative information and so is typically used for factor screening \cite{Saltelli2008,Saltelli2002}. However, being a semi-quantitative method, there is no definitive boundary separating the important and unidentifiable parameters. It turns out that in most models with a large number of variables ($>20$), there are only a few influential variables with many uninfluential ones in between \cite{Saltelli2008}. Hence, in practice, a line of demarcation separating influential and less important variables can often be ascertained qualitatively. Nonetheless, the Morris method is at a disadvantage when dealing with factor interactions. Though able to detect if a factor is involved in nonlinearities or interactions, it cannot determine which of these is present, nor, in the case of interactions, can it identify which factors are involved. Instead it gives only one lumped measure, $\sigma_i$, of the total magnitude of its interactions and nonlinearities. In situations where more clarity is required, we turn to the more computationally expensive variance-based methods discussed next.

Being a popular method, the Morris method has been used extensively for SA of models in the biomedical sciences \cite{Laranjeira2017, Laranjeira2018,Hall2015}, hydrology \cite{Wainwright2014a} and chemical engineering \cite{Sin2009}, among other applications.

\medskip

\begin{example}\upshape
To implement the Morris method on model \eqref{ToyModel}, we set the number of elementary effects per factor to be 100 and choose the step-size $\Delta=1/100$. The variables $X_1$ and $X_4$ have the largest values of $\mu^*$ and so are the two factors that most affect the output variance. The other two factors, $X_2$ and $X_3$, have less effect on the output, having lower values of $\mu^*$. We now turn our attention to the magnitude of interactions and nonlinearities that each factor is involved in, represented by the $\sigma$ values. With the exception of $X_4$, the values of $\sigma_i$ for all other factors are nonzero, since the only linear term in model \eqref{ToyModel} is the term $X_4$. We note that the Morris sensitivity indices of the first three factors will vary unless $\Delta$ is chosen to be sufficeintly small. However, the indices of $X_4$ stays the same regardless of the step size, again due to its appearance only as a linear term. 
Finally, since $X_2$ and $X_3$ interact with each other, their $\sigma_i$ values are similar.

As noted previously, one feature of the Morris method is that it allows us to find any unidentifiable factors (having little or not influence on the output). In our example, however, since the $\mu^*$ of each variable is much greater than $0$, we conclude that all the factors play a role in the output variance.  
\end{example}

\begin{table}
\begin{center}
\begin{tabu}{lll}
\tabucline[1pt] {-}
\hline
Factor & \multicolumn{1}{c}{$\mu^*_i$} & \multicolumn{1}{c}{\begin{tabular}[c]{@{}c@{}} $\sigma_i$ \\ \end{tabular}} \\
\hline
$X_1$  & 1.04                        & 0.58                                                                                          \\
$X_2$  & 0.51                        & 0.28                                                                                          \\
$X_3$  & 0.47                        & 0.30                                                                                           \\
$X_4$  & 1.00                        & 0.00                                                                                              \\
\hline
\tabucline[1pt] {-}
\end{tabu}
\caption{The results obtained upon implementing the Morris method on model \eqref{ToyModel}. We show 
$\mu_i^*$ and $\sigma_i$ when the number of runs, $r$, is set to 100, and the step size $\Delta=1/100$. }
\label{TableMorrisToy}
\end{center}
\end{table}

%---------------------------------------------------
\subsection{Sobol method}\label{sobol}
In the following two sections, we will discuss two common variance-based SA techniques. These methods calculate the proportion of the model variance caused by each input factor, as well as its interaction with every other factor. The trade-off, however, is that variance-based methods are computationally expensive.

The first of these variance-based techniques is the Sobol method \cite{Sobol1993, Sobol2001}, which decomposes the model variance into contributions from each factor as well as all its interactions. The Sobol sensitivity indices can be written in terms of conditional variances \cite{Jansen1999, Oakley2004, Saltelli2010}. For instance, the {\it first-order indices},  or {\it main effects}, measure the direct contribution of each input factor to the output variance
\begin{equation}\label{SobolFirstIndices}
S_i=\frac{\V_{X_i}(\E_{X_{\sim i}}(Y| X_i))}{\V(Y)}, % = \frac{\V(Y)-\E_{X_i}(\V_{X_{\sim i}}(Y|X_i))}{\V(Y)},
\end{equation}
where $\E$ is the expected value, $\V$ the variance, $X_{\sim i}$ denotes all model factors except the $i$th and $Y$ is the model output. For the expectation operator, the mean of $Y$ is taken over all possible values of $X_{\sim i}$ (denoted in the subscript) while keeping the factor $X_i$ fixed. The variance, on the other hand, is taken over all possible values of $X_i$.  The {\it total-order indices}, or {\it total effects}, introduced by Homma and Saltelli \cite{Homma1996}, measure the part of output variance explained by all the effects in which it plays a part and is defined as
\begin{equation}\label{Sobol_total}
S_{T_i}=\frac{\E_{X_{\sim i}}(\V_{X_i}(Y|X_{\sim i}))}{\V(Y)}.
\end{equation}
The first-order indices $S_i$ and total indices $S_{T_i}$ can be interpreted in terms of expected reduction of variance \cite{Saltelli2010}. Note that $\V_{X_i}(\E_{X_{\sim i}}(Y| X_i))$ can be thought of as the expected reduction in model variance that would be obtained if $X_i$ was to be fixed. Also, $\E_{X_{\sim i}}(\V_{X_i}(Y|X_{\sim i}))$ is the expected variance that would be left if all factors but $X_i$ were to be fixed. 

If the total-order index of a factor is zero, then that factor is non-influential. Thus, $S_{T_i}$ is especially well-suited for factor screening. First-order indices, on the other hand, are typically used for ranking, especially when the interactions with other variables do not contribute significantly to the output variance \cite{Pianosi2016}. It is also possible to define sensitivity indices of any order. For example, second-order indices measure the contribution made by the interaction between a pair of factors to the output variance and so on \cite{Borgonovo2007}. 

There are two main steps involved in the computation of Sobol indices. The first is to draw sample values for the input factors. Three appropriate sampling methods are discussed in Section \ref{samplingsection}. Other sampling schemes originally designed for the computation of variance-based sensitivity indices include the winding stairs method \cite{Jansen1994}, which uses either pseudorandom \cite{Chan2000} or quasi-random sequences \cite{Owen1998}, to generate values for each factor and the radial design \cite{Saltelli2002}, which was shown in \cite{Saltelli2010} to be the more efficient of these two schemes for calculating the total-order Sobol indices of several test functions. 

 The second step is to calculate the sensitivity indices, $S_i$ and $S_{T_i}$, using the samples generated from the sampling scheme. Several algorithms have been introduced for this purpose. The available algorithms for $S_i$ include Sobol's approach \cite{Sobol1993} as well as later versions by Jansen \cite{Jansen1999} and Saltelli et al. \cite{Saltelli2010}, among others. There are also many algorithms for implementing $S_{T_i}$. Homma and Saltelli, for example, not only introduced this index but also gave the first algorithm for its calculation \cite{Homma1996}. Other versions by Jansen \cite{Jansen1999} and Sobol \cite{Sobol2007} were proposed later. Of these three estimators of $S_{T_i}$, Jansen's algorithm was shown to be the most efficient \cite{Saltelli2010}.

Several extensions to the Sobol method have been proposed including the case of dependent variables \cite{Kucherenko2012} as well as stochastic and statistical models \cite{Hart2017, Hart2019b}. It has been used extensively to analyse models in, for example, the fields of hydrology \cite{Nossent2011,Zhang2013} and biomedical science \cite{Ojeda2016, Jarrett2016}, including analysis on high-dimensional models \cite{Hart2019a}. 

\begin{table}[]
\begin{center}
\begin{tabu}{llcclcc}
\tabucline[0.7pt] {-}
\multirow{2}{*}{Factor} 	&\qquad& \multicolumn{2}{c}{Sobol} 		&\qquad					& \multicolumn{2}{c}{FAST/eFAST} \\
                        			&& $S_i$      				& $S_{T_i}$    		&& $S_i$       					& $S_{T_i}$     \\
\hline
$X_1$                   		&& 0.403			   		& 0.403        		&& 0.403				        		& 0.408         \\
$X_2$                   		&& 0.094     				& 0.126        		&& 0.094      					& 0.131         \\
$X_3$                   		&& 0.094     				& 0.126        		&& 0.094      					& 0.131        				 \\
$X_4$                   		&& 0.377     				& 0.377        		&& 0.368       		& 0.373        \\
\tabucline[0.7pt] {-}
\end{tabu}
\caption{This table shows the first and total-order sensitivity indices of each factor in the model \eqref{ToyModel}, calculated using both the Sobol and FAST/eFAST methods.}
\label{VarianceBasedTable}
\end{center}
\end{table}

\medskip

\begin{example}\upshape
The Sobol indices  for model \eqref{ToyModel} are given in Table \ref{VarianceBasedTable}. 
Note that $X_1$ is the most sensitive factor; and $X_2$ and $X_3$ are the only variables involved in interactions. Note how this differs from the Morris method, where $\sigma_i$ is non-zero for not only these two factors but $X_1$ as well. Hence, running both the Morris and Sobol methods in tandem and subsequently comparing $\sigma_i$ with $S_{T_i}-S_i$ isolates the nonlinear and interaction (or {\it additional}) effects of each factor, as noted by Wainwright et al. \cite{Wainwright2014a}. We also illustrate this through an example (see Section \ref{ODECellDiffModel}).
\end{example}

%---------------------------------------------------
\subsection{FAST and eFAST}  \label{Sec:FAST}
Introduced by Cukier et al. \cite{Cukier1973}, the {\it Fourier Amplitude Sensitivity Test} (FAST) is another variance-based method. The FAST method uses sinusoidal functions and the orthogonality property of the Fourier series to give an approximation of the total model variance in terms of the real and imaginary coefficients of the Fourier series. Then, the first-order index of factor $X_i$ is given by the proportion of this total variance attributable to the Fourier series harmonics caused by $X_i$. One of the first numerical implementations of this was developed in \cite{McRae1982}. 
However, in this form, the FAST method can calculate only the first order indices of each factor. Saltelli et al. \cite{Saltelli1999} extended the method to include the computation of the total indices as well, giving rise to the {\it extended FAST} (eFAST).

Being a variance-based method, FAST/eFAST gives quantitative information, contained in the first and total-order sensitivity indices. The eFAST is also more efficient than the Sobol method as the former calculates all indices in one set of model evaluations [10]. However, all these methods suffer from being more computationally expensive than the derivative-based methods as well as the correlation and regression methods of the next section.

The FAST methods have been commonly applied to analyse models in engineering and science: a set of three papers by Cukier et al. \cite{Cukier1973,Cukier1975} and Schaiby et al. \cite{Schaibly1973} investigating chemical kinetic systems introduced the FAST method and, since then, using  eFAST, researchers have expanded its areas of application to include, for instance, SA of reliability/risk models including hydrological models assessing flooding risk \cite{Crosetto2000}, modelling of atopic dermatitis \cite{Tanaka2011} and thermodynamic models for gene transcription \cite{Dresch2010}.

\medskip

\begin{example}\upshape
The results of the FAST/eFAST method on model \eqref{ToyModel}  are given in Table \ref{VarianceBasedTable}. Note, as expected, the similarities between the results of Sobol and FAST/eFAST. 
\end{example}

%---------------------------------------------------
\subsection{DGSM}  \label{Sec:DGSM}
Derivative-based global sensitivity measures (DGSM) can be thought of as an extension of the local sensitivity and Morris methods, as it involves taking the average of the partial derivative of the model with respect to the input factor across factor space  \cite{Sobol1995, Kucherenko2009, Sobol2010}. Consider a differentiable function $f(x_1,\ldots,x_n)$ defined in the unit hypercube $H^n = [0,1]^n$. It has been shown \cite{Sobol2009} that the modified Morris measure $\mu^*$  (see equation \eqref{MorrisMethodMuStar}) is an approximation of the functional $\int_{H^n}|{\p f}/{\p x_i}|dx$. The DGSM measures are defined  as
\begin{equation}
v_i = \int_{H^n}\Big(\frac{\p f}{\p x_i}\Big)^2 dx.
\end{equation}

The calculation of DGSM indices is typically performed using Monte Carlo, quasi-Monte Carlo or Latin hypercube sampling \cite{Touzani2014}. The use of Sobol sequences or other low-discrepancy number generators may improve the computational efficiency \cite{Kucherenko2009}.
An improvement to the method's numerical efficiency using automatic differentiation has also been considered \cite{KiparissidesKucherenko2009}.

Also found was a link between the DGSM and the total-order sensitivity index, $S_{T_i}$, of the Sobol method. In the cases of uniformly and normally distributed factors, it can be shown that $S_{T_i}\leq {v_i}/{\pi^2\, \V(Y)}$, where $ \V(Y)$ is the total model variance \cite{Sobol2010}.

The method can be extended for use on both single and groups of inputs \cite{Sobol2010a}. The DGSM indices give similar rankings to those based on the Sobol method if the model is linear or quasilinear but for highly nonlinear models the rankings given by the two methods may no longer agree \cite{Kucherenko2009}. 
 Applications of DGSM in the fields of hydrology \cite{Touzani2014} and biochemistry \cite{Kiparissides2009} are available, as well as a survey article investigating its link to Sobol sensitivity indices in \cite{KucherenkoIoos2015}.  
 
\medskip 
 
\begin{example}\upshape
 We applied the DGSM methods to model \eqref{ToyModel}, obtaining the results shown in Table \ref{TableDGSMToy}. Note that rescaling the indices by the factor $\pi^2 \V(Y)=2.18$ produces an upper bound for the Sobol indices (Tables \ref{VarianceBasedTable} and \ref{TableDGSMToy}) as shown in \cite{Sobol2010}. The factor rankings remain the same as those given by the Sobol method (Table \ref{VarianceBasedTable}).
\end{example}

\begin{table}
\begin{center}
\begin{tabu}{lcc}
\tabucline[1pt] {-}
\hline
Factor   & $v_i$ & $v_i/(\pi^2 \V(Y))$   \\
\hline
$X_1$ & 1.33  & 0.61                     \\
$X_2$ & 0.33  & 0.15                     \\
$X_3$ & 0.33  & 0.15                     \\
$X_4$ & 1.00  & 0.46                     \\
\hline
\tabucline[1pt] {-}
\end{tabu}
\caption{The results of the DGSM index $v_i$ analysis and the rescaled version $v_i/(\pi^2 \V(Y))$ on model \eqref{ToyModel}.}
\label{TableDGSMToy}
\end{center}
\end{table}

%---------------------------------------------------
\subsection{Sensitivity index}  \label{Sec:SI}
The {\it sensitivity index} ($\SI_i$) computes the percentage difference when varying one factor within its typical range (e.g. from maximum to minimum value):
\begin{equation}
\label{SenseIndex}
\SI_i=\frac{D_{i,\max}-D_{i,\min}}{D_{i,\max}},
\end{equation}
where $D_{i,\max}$ and $D_{i,\min}$ are the maximum and minimum values of the output, respectively, when varying the independent variable $X_i$ and keeping the other variables, constant typically set to their baseline values \cite{Helton1993}. 

\medskip

\begin{example}\upshape
Calculation of the sensitivity index for each factor of the model \eqref{ToyModel}, assuming all factors are initially set to 0.5, gives 
$\SI_1=0.57$, $\SI_2=\SI_3=0.4$ and $\SI_4=0.67$. The most sensitive parameter here is $X_4$, however, the ranking depends on the initial factor values, just as is the case with other local methods. 
\end{example}

%---------------------------------------------------
\subsection{Importance index}  \label{Sec:Importance}
The {\it importance index} ($I_i$) computes the ratio between the variances of a factor $X_i$ and output $Y$ \cite{Hamby1994}:
\begin{equation}
\label{ImportanceInd}
\I_i=\frac{\sigma^2_{X_i}}{\sigma^2_{Y}}.
\end{equation}
This is a simple sensitivity technique that can be thought of as a measure of a fractional contribution to the total variability of the model. Since $\sigma^2_{X_i}$ is a measure that takes into account the entire range of the $i$th input, we consider the $I_i$ to be a global method. 

\medskip
\begin{example}\upshape
The importance indices computed for model \eqref{ToyModel} are the samea $\I_1=\I_2=\I_3=\I_4={1}/{14}$ as each factor follows a uniform distribution between 0 to 1. 
\end{example}

%-------------------------------
\subsection{Correlation and regression methods} \label{corregress}

The underlying principle of the correlation and regression methods is to obtain measures of a factor's sensitivity based on the extent of its correlation with the output. There are several closely-related methods here, each of which can be calculated with a set of input data points and the associated output. The distribution of each factor is first determined, before samples are generated using one of several potential sampling methods. Of these, simple random sampling (see Section \ref{samplingsection}) works well for computationally cheap models, where many samples can be generated. As the computational cost to evaluate the model increases, however, the Latin hypercube method (see Section \ref{samplingsection}) becomes a more viable choice \cite{Helton2006}. Other methods can also be used situationally. For instance, Importance Sampling takes into account regions of factor space that have {low probability but high consequence} \cite{Helton2006}. 
The model output is then evaluated using these samples, before the correlation methods discussed below give sensitivity measures for each factor \cite{Helton2006}. 

 We introduce five such methods, the first three of which give a measure of the linear correlation between the input and output, whereas the remaining two examine any monotonic relationship between them. All methods give a measure of the factor's importance and so can be used for factor ranking \cite{Kleijnen1999}. Note that one must first verify that the input and output relationship is, in fact, linear or monotonic before applying these methods \cite{Schober2018}. Failure to do so may lead to a misleading analysis, as the model may not satisfy these conditions. 

\medskip

\noindent
{\bf Pearson's correlation coefficient.} 
The {\it Pearson's correlation coefficients} ($\CC_i$), is a measure of the linear relationship between the input factor $X_i$ and output $Y$ and is computed as follows
\begin{equation}
\label{CCindex}
\CC_i=\frac{\sum_{j=1}^{n}(X_{ij}-\bar{X_i})(Y_j-\bar{Y})}{{[ \sum_{j=1}^{n} (X_{ij}-\bar{X_i})^2 \sum_{j=1}^{n}(Y_j-\bar{Y})^2]}^{\frac{1}{2}}},
\end{equation}
where $\bar{X_i}$ and $\bar{Y}$ are the corresponding sample means;  $X_{ij}$ and $Y_j$ are the $j$th sample point of the input $X_i$ and the output $Y$, respectively, and $n$ is the sample size.

The coefficients $\CC_i$ take values between $-1$ and $1$, with both values denoting a perfectly linear relationship between input $X_i$ and output $Y$;  a value of $0$ implies that there is no linear correlation between the input and output data. 

\medskip

\noindent
{\bf Standardised regression coefficient.} There is a class of methods for computing the sensitivity of linear models for which the input and output can be modelled as
 \begin{equation}\label{SRCinout}
\hat Y=b_0+\sum_{i=1}^{k}b_iX_i,
\end{equation}
where $b_0, b_1,\ldots, b_k$ are the coefficients determine through the process of model calibration {\cite{Rao1998,Seber2003,Myers1990,Bland2015}}. 
To remove the influence of units in which the factors are measured, we normalise the variables. Instead of regressing $Y$ on $X_i$'s, we regress $(\hat Y-\bar Y)/\hat\sigma_{Y}$ on ${(X_i-\bar{X}_i})/{\hat\sigma_{X_i}}$, where $\bar{X_i}$ and $\bar{Y}$ are the sample means and  $\hat\sigma_{X_i}$ and $\hat\sigma_{Y}$ are the sample standard deviation of the corresponding variables. 
The {\it standardised regression coefficients} ($\SRC_i$) are the resulting normalised linear regression coefficients
\begin{equation}
\label{SRC}
\SRC_i=b_i \frac{\hat\sigma_{X_i}}{\hat\sigma_{Y}}.
\end{equation}
The ordering of the absolute values of these unit-free coefficients can be used to rank the relative importance of the factors $X_i$. 

\medskip

\noindent
{\bf Partial correlation coefficient.} If the correlation between independent variables exists, it may strongly bias the values of the Pearson's correlation coefficients. The {\it partial correlation coefficient} ($\PCC_i$) accounts for this by measuring the strength of linear relationship between $X_i$ and $Y$ after both variables have been adjusted for all the remaining independent variables. It is computed as
\begin{equation}\label{SRC}
\PCC_i=\SRC_i \sqrt{\frac{1-R^2_{X_i}}{1-R^2}},
\end{equation}
where $R^2$ is the coefficient of determination and $R_{X_i}^2$ is the coefficient of determination obtained from regressing $X_i$ on $Y$ and $X_{\sim i}$ (a vector of independent variables or factors except $X_i$).  

$\PCC_i$ can be interpreted as a measure of the new information gained from introducing the $i$th factor to the model \cite{Rao1998}. If all input factors and the output have been standardised to mean of 0 and variance of 1, then the $\SRC_i$ is equivalent to $\PCC_i$ \cite{Kleijnen1999}. The $\PCC_i$ can also be calculated using two other methods. The first involves finding the Pearson's correlation of the residuals for the two models found by regressing, first, $X_{\sim i}$ on $Y$ and, second, $X_{\sim i}$ on $X_i$. The other method requires the calculation of the inverse of the correlation matrix, as shown in \cite{Iman1984}.

\begin{table}[]
\begin{center}
\begin{tabu}{lccccc}
\tabucline[0.7pt] {-}
Factor & \multicolumn{1}{c}{$\CC_i$} & \multicolumn{1}{c}{$\SRC_i$} & \multicolumn{1}{c}{$\PCC_i$} & $\SRCC_i$ & $\PRCC_i$ \\ \hline
$X_1$  & 0.61                                                        & 0.49                                                           & 0.93                                                       & 0.60                             & 0.91                             \\ 
$X_2$  & 0.30                                                        & 0.19                                                           & 0.79                                                       & 0.29                       	 & 0.73                            \\
$X_3$  & 0.31                                                        & 0.19                                                           & 0.79                                                       & 0.29                             & 0.72                            \\
$X_4$  & 0.62                                                        & 0.50                                                           & 0.93                                                       & 0.62                             & 0.91                            \\
\tabucline[0.7pt] {-}
\end{tabu}
\caption{This table shows the sensitivity indices of each factor in model \eqref{ToyModel}, calculated using various regression methods.}
\label{RegressTable}
\end{center}
\end{table}

\medskip

\noindent
{\bf Spearman and partial rank correlation coefficients.} While both the Pearson's and partial correlation coefficients quantify the extent of linearities in the data, the Spearman ($\SRCC_i$) and partial rank correlation coefficients ($\PRCC_i$) are their respective counterparts for nonlinear models. These indices quantify the monotonicity inherent in the data by linearising them, using their numerical ranking \cite{Kleijnen1999}.

Specifically, $\SRCC_i$ and $\PRCC_i$ are calculated using the formulae for $\CC_i$ and $\PCC_i$, respectively, but replacing $X_i$ and $Y$ with the rank values of those variables. Hence $\SRCC_i$ can be seen as a measure of the linearity of each factor's ranked data points, while $\PRCC_i$ accounts for any correlations between the factors \cite{Kleijnen1999}. An index of $1$ or $-1$ indicates a monotonically increasing or decreasing function, respectively, whereas an index of $0$ occurs if successive data points increase then decrease alternately. The $\PRCC_i$ has been applied, for instance, to models of HIV transmission \cite{Blower1994}.

\medskip 

\begin{example}\upshape
Table \ref{RegressTable} shows the results when the five regression methods mentioned above are implemented on model  \eqref{ToyModel}. As is consistent with other sensitivity methods, all the results here suggest that the most influential parameters are $X_1$ and $X_4$. It is noticeable that the values of the Pearson $\CC_i$ and Spearman $\SRCC_i$ are similar, as well as those of the partial $\PCC_i$ and partial ranked $\PRCC_i$ correlation coefficients. This is not generally the case. In our model, the four input variables form three terms which are added together to make a monotonic function. 

However, the Spearman coefficients are not $1$ because of the presence of the other terms. Each variable must `compete' with the others for impact on the output, $Y$, and this competition reduces the impact of an increase in any individual variable alone. This is particularly the case for $X_2$ and $X_3$, where an increase in one variable may not even produce an increase in the $X_2X_3$ term, if the other variable is sufficiently decreased; this is reflected in the lower Spearman coefficients of these two variables. If we were to increase the $X_1^2$ term by a factor of $10$, for instance, then $\PCC_1$ and $\SRCC_1$ increase to around $0.96$ and $1.00$, respectively, as now $10X_1^2$ is the term dominating the others, with respect to the output.

Although Pearson's correlation is a measure of linearity, when the input values are positive, as they are here, the nonlinear term, $X_1^2$, nonetheless has some resemblance to a linearly increasing function, such that $\CC_1$ is still relatively high. What is interesting is that $\CC_1$ and $\PCC_1$ decrease almost to zero when the first term of the model, $X_1^2$, is modified to $(X_1-0.5)^2$. After this shifting of $X_1$, the quadratic term is no longer monotonic between 0 and 1; indeed, about half the points should sit on either side of the arms of the parabola, leading to the large decrease in both correlation coefficients. Such an example shows a possible limitation in both these methods for sensitivity analysis: they require model linearity (Pearson) or monotonicity (Spearman), and even a slight change to the model that negates these conditions will result in very different correlation scores.  
Similar arguments can be applied to explain the partial correlation coefficients. 
\end{example}

%---------------------------------------------------
\subsection{Sensitivities in the ODE setting}
Models in biomedical sciences often appear in the form of Ordinary Differential Equations (ODEs), which can be written as
\begin{eqnarray}
 \frac{d x}{dt}&=&f(t,x,\theta) \label{ode:mod},\qquad x(t_0)=x_0,
\end{eqnarray}
where $x=[x_1,\ldots,x_n]\in\R^n$  denotes the state vector, $t$ is time and $\theta=[\theta_1,\theta_2,\ldots,\theta_p]\in\R^p$ the parameter vector.
Here the factors are taken to be the parameters $\theta$, as we aim to quantify how much they influence the variability of the model output $x(t)$ \cite{Banks2007, Banks2007a}. Therefore the {\it sensitivity functions}, which describe the time evolution of the dependence of model output on model parameters, are defined as
\begin{equation}\label{ode:sens}
S_{\theta_j}^{x_i}(t) = \frac{\partial x_i(t)}{\partial \theta_j},
\end{equation}
where $j=1,\ldots,p$ are the parameter indices and $i=1,\ldots,n$ are the state variable indices. We note that for an ODE model with $p$ parameters and $n$ state variables there will be $n\times p$ sensitivity functions. 

A SA of time-dependent ODE models can be conducted in two different ways. The first possibility is to calculate the sensitivity at specific time-points, if we are interested only in those particular instances \cite{Laranjeira2017, Laranjeira2018, Tanaka2011}. The second possibility is to calculate the sensitivity over the entire interval on which the model is defined \cite{Wainwright2014a,Wu2013,Bianca2012}.

The methods reviewed in Section 3 can be applied to ODE models by first solving the model equations and then calculating the sensitivity at specific time points. There are also methods designed specifically for ODE models; these are reviewed below. 

\medskip

\noindent{\bf Finite difference method.} The finite difference method \cite{Iott1985,Haftka1989,Pope2009} for calculating the sensitivities \eqref{ode:sens} approximates them using the forward differences
\[
\frac{\partial x}{\partial \theta_j} \approx \frac{x(t,\theta+he_j)-x(t,\theta)}{h}, \qquad
e_i = \Big[0 . . . 0 \, \overset{j}{\widehat{1}} \, 0 . . . 0\Big],
\]
where $e_j$ is the unit vector in the $j$th component direction. To compute the sensitivities  \eqref{ode:sens} for each of $p$ parameters, we require one solution of the model with baseline values of parameters $\bar\theta_j$ and $p$ solutions of the model with perturbed parameters $\bar\theta_j+\Delta \bar \theta_j$ for $j=1,\ldots,p$. We note here that this is an OAT method, where one parameter is perturbed at a time while the rest are kept constant at the nominal value (see Section \ref{sec:OATMAT}). 

\medskip

\noindent{\bf Direct method.}
The direct method \cite{Atherton1975,Dickinson1976} relies on deriving the {\it sensitivity equations} obtained by differentiating model \eqref{ode:mod} with respect to the parameter $\theta_j$, i.e.
\begin{equation}\label{ode:seneq}
\frac{d}{dt} \Big(\frac{\partial x}{\partial \theta_j}\Big)=\frac{\partial f}{\partial x}\frac{\partial x}{\partial \theta_j}+\frac{\partial f}{\partial \theta_j},
 %       \frac{d}{dt} S^{x}_{\theta_j} &= J_{x}\cdot S^{x}_{\theta_j}+J_{\theta_j}
 \end{equation}
where ${\partial f}/{\partial x}$ is the model's Jacobian matrix (here assumed to be obtained explicitly); and ${\partial f}/{\partial \theta_j}$ is a vector of derivatives with respect to $\theta_j$.  Note that to compute the sensitivity $S^x_{\theta_j}(t) = \partial f(t,x,\theta)/\partial\theta_j$ from the sensitivity equation \eqref{ode:seneq}, we need knowledge of $x(t)$ at each point. Thus we obtain $S^{x}_{\theta_j}$ by simultaneously solving model equation  \eqref{ode:mod} and the sensitivity equations \eqref{ode:seneq} with the initial $S^x_{\theta_j}(0) = 0_{n\times p}$, here $0_{n\times p}$ is the $n\times p$ zero matrix (obtained by differentiation with respect to $\theta$, with the initial condition $x(0)=x_0$). This approach has been found to be unstable for certain stiff kinetic problems. The sensitivities are computed by solving a large system of ODEs resulting from coupling the model with an auxiliary sensitivity equations and therefore the method is computationally suboptimal. 

\medskip

\noindent{\bf Green function method}.
The Green function method for solving ODEs applied to SA is considered in \cite{Hwang1978}. It calculates the Green function for the auxiliary equations and produces the sensitivity coefficients from integrals over the Green function, which is the integral kernel of the solution of ODEs. There are a variety of these methods, including the popular {\it Analytically Integrated Magnus} (AIM) method \cite{Kramer1981}.

\medskip

\noindent{\bf Automatic differentiation method.}
The automatic differentiation method also has been applied to compute model sensitivities \cite{Hwang1997}. It allows us to compute the derivatives of any arbitrary order by applying the chain rule \cite{Griewank1989}. Computer routines have been developed for this purpose in different programming languages, as well as packages such as ACADO \cite{Hwang1997} and CasADi \cite{Andersson2012}. 

\noindent{\bf Parameter estimation of biological models.}
As part of model validation one needs to estimate model parameters from some data measurements.
However, biomedical data are often either difficult to obtain (e.g. invasive methods may be necessary) or costly. Therefore, it might be desirable to optimise the data collection process (e.g. by minimising the intervention while preserving the level of information). In particular, it is important to determine at which time points the measurements are most informative for estimating a given model parameter and how the process depends on the number of sampled data points \cite{Capaldi2012}. SA is one way to address this problem and has been performed in the context of asymptotic statistical theory \cite{Seber1989, Davidian1996}.  Example applications of this theory to epidemiological models are available \cite{Banks2009a, Banks2009, Cintron-Arias2009}.

Since the sensitivities of a time-dependent system provide temporal information on how the states of a system vary to changes in the parameters, they can be used to determine the time intervals during which the system is the most sensitive to such changes. The key link here is that the asymptotic statistical theory allows us to calculate standard errors of parameter estimates through the use of sensitivity functions which, in turn, provide information about the time periods during which the data points carry the most information about the estimation process \cite{Banks2007a, Banks2007}. Therefore, if the sensitivity function $S^{x_i}_{\theta_j}(t)$ is close to zero, in some time interval, changes in the parameter $\theta_j$ will have little effect on the state variable $x_i(t)$.

%%%%%%%%%%%%%%%%%%%%%%%%%%%%%%%%%%%%%%%%%
\section{Visual methods} \label{GraphicalSection}
SA methods can also be categorised by how the results are presented. Visual methods are known as {\it qualitative}, while those that give numerical values representing the sensitivity indices are {\it quantitative}. Though qualitative, visual tools can enable an intuitive understanding of the important factors.

\begin{figure}[t!]
\centering
\includegraphics[width=\linewidth]{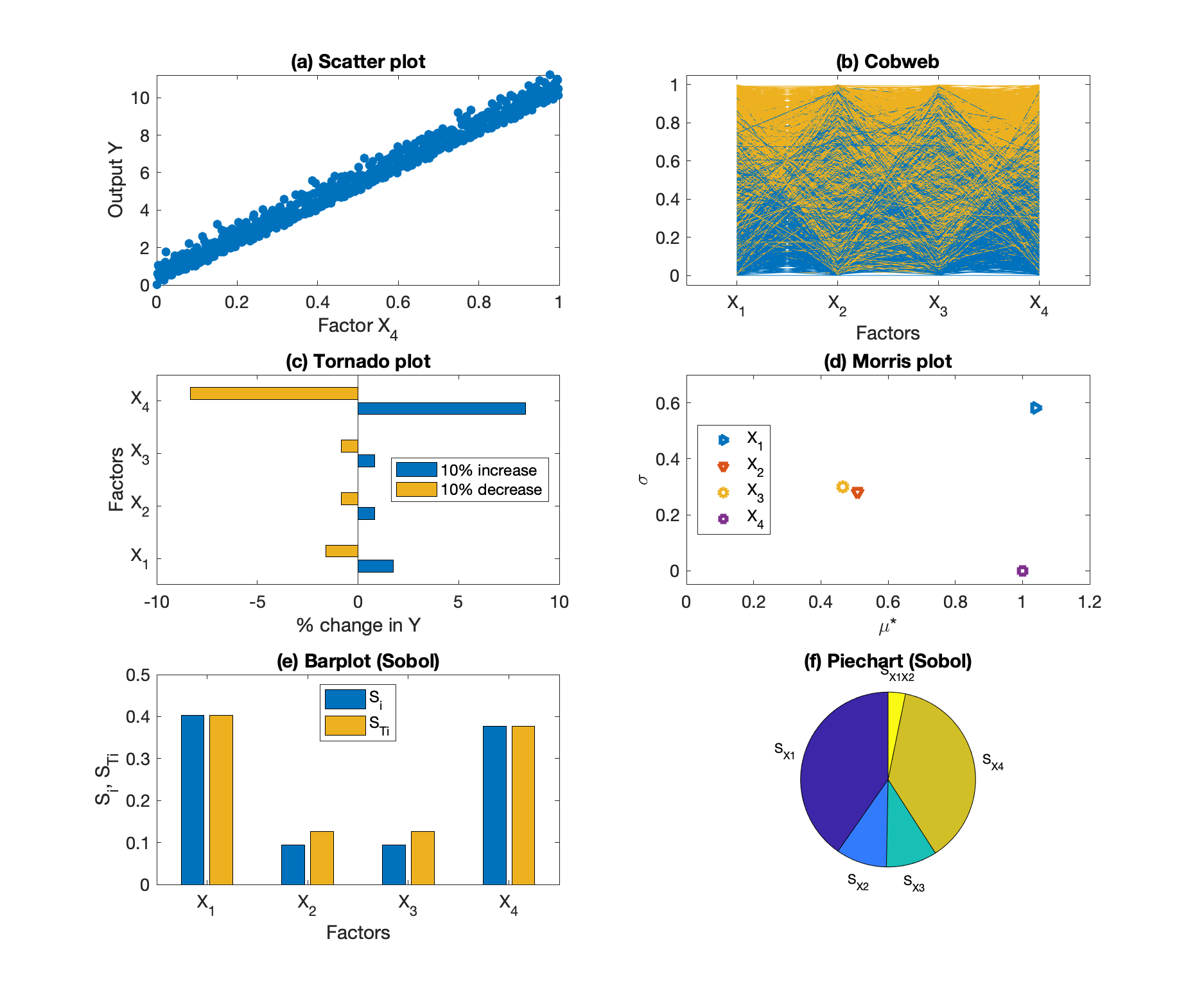}
\caption{Visualisation methods for input-output (top row) and SA indices (middle and bottom rows).}\label{fig:all_sa}
\end{figure}

%-----------------------------------------
% Input-Output visualisation
%-----------------------------------------

\smallskip
 
\noindent
{\bf Scatterplot.} The most common way to visualise the input-output relationship is by plotting the model output against a specific input \cite{Kleijnen1999, Saltelli2000}. Scatterplots can be useful for a general understanding of the magnitude of the underlying sensitivities as well as detecting some functional relation between model input and output including nonlinearities, albeit in a subjective manner. The inspection of scatterplots is usually a good starting point for performing sensitivity analysis \cite{Helton1993}. The scatterplot in Figure\,\ref{fig:all_sa}a shows the relationship between the variable $X_4$ and $Y$ for the example model \eqref{ToyModel}. 

\smallskip

\noindent 
 {\bf Cobweb plot.} Another way to visualise the input-output relationship is by plotting the distribution of all the input samples: these are known as either cobwebs or parallel coordinate plots \cite{Saltelli2000, Iooss2015, Pianosi2016}. Here, the range of each of the $n$ input factors is represented by a vertical line, or axis. Then, taking the first set of input samples, each of the $n$ points of the set is plotted on its appropriate axis, before successive points are connected by lines.  

The process is repeated for the remaining sets of input samples, creating a plot resembling a cobweb. We can also impose certain properties on a specific input factor to find the values the other factors should take \cite{Saltelli2000}. For instance, we may be interested in keeping one variable over some threshold value and so require information on the values of the remaining factors. In this case, we discard the lines where the threshold is not exceeded and examine the resultant graph, known as a {conditional cobweb plot} \cite{Saltelli2000}. Figure\,\ref{fig:all_sa}b shows a cobweb plot with the distribution of all samples of the four input variables for the example model \eqref{ToyModel}.

%-----------------------------------------
% SA visualisation
%-----------------------------------------

\bigskip

While the scatterplot and cobweb are tools for input-output visualisation, the following four methods are commonly used to visualise sensitivity indices. 
 
\smallskip
 
\noindent
 {\bf Tornado plot.}  This is a bar graph that is typically used to illustrate one-way sensitivities (see Section\,\ref{Sec:oneway}). A tornado plot shows percentage change in the output after one factor is increased or decreased by a certain percentage (e.g. 10\%) when all others are fixed to that nominal value  \cite{Howard1988, Saltelli2000}.  The tornado plot in Figure\,\ref{fig:all_sa}c shows the one-way sensitivities of the four input variables in the example model \eqref{ToyModel} when we increase and decrease their values by 10\% from their baseline.
 
\smallskip
 
\noindent
{\bf Morris plot.} The Morris method (see Section\,\ref{Sec:morris}) has its own graphical representation. The mean of absolute value of the elementary effects $\mu^*$ is plotted against the standard deviation of the elementary effects $\sigma$. This gives a visual representation of the factor ranking ($\mu$ axis) and their interactions with other variables ($\sigma$ axis).  The bottom-left corner of the graph indicates little effect on the output ($\mu^*$ small) and few interactions with other factors ($\sigma^*$ small). The top-right corner indicates greater effect on model output ($\mu^*$ big) and existence of interactions with other factors  ($\sigma^*$ large) \cite{Saltelli2008}. The Morris plot in Figure\,\ref{fig:all_sa}d illustrates the $\mu^*$ and $\sigma$ sensitivity values for the four factors of the example model \eqref{ToyModel}.

\smallskip

\noindent
{\bf Bar plots}. This is a common way to graphically represent different sensitivity indices, where each measure is plotted as a separate bar. Figure\,\ref{fig:all_sa}e shows a double barplot for the first-order ($S_i$) and total-order Sobol sensitivity indices ($S_{T_i}$) corresponding to four input variables in the example model \eqref{ToyModel}. Note that the barplot is a convenient way to illustrate the results of the sentivity analysis, but these can also be presented in a table (Table\,\ref{VarianceBasedTable}).

\smallskip

\noindent
{\bf Pie chart}. This is another popular way to present the results of a variance-based SA \cite{Marino2008a}. Each index is represented by a slice of the pie proportional to its magnitude. Figure\,\ref{fig:all_sa}f shows a pie chart illustrating how the sensitivity indices of the example model \eqref{ToyModel} can be broken down into its constituent Sobol indices.

\smallskip

\noindent
{\bf Radar graphs}. These are generally less common in biomedical applications but can portray the same information \cite{Saltelli2000}. The advantage a radar graph has over tornado and bar plots is the ability to concisely summarise information from models with a large number of variables (see \cite[Figure 11.17]{Saltelli2000}). As shown in that example, a radar graph represents each factor as a ray extending from the origin. Each factor's sensitivity is plotted at the appropriate point on the ray and connected to its two adjacent sensitivities by a straight line.

\smallskip 

\noindent
 {\bf Circular diagram}. Also known as {\it radial convergence plot}, these can be used for variance-based methods illustrating the first and total-order sensitivities of each factor as smaller and larger circles, respectively. The size of these circles is proportional to the magnitude of the sensitivity. A line between two factors indicates an interaction; the width of the line is proportional to the extent of this interaction, see \cite{Butler2014}.

\smallskip 

\noindent
{\bf Matrix plots}. These plots extend the concept of the scatterplot to multivariate models by presenting a grid of scatterplots. Each variable is found along both the horizontal and vertical axes, such that the scatterplot located in the first row and $n$th column shows the relationship between the first and $n$th variable \cite{VanWerkhoven2008}. Such a visual representation faciliates the observation of trends such as the correlation between two variables as well as any outliers.

The sensitivities of the parameters in an ODE model can be calculated at any point in time at which the model is defined. While often the sensitivities are required only at a certain point in time (e.g. the steady-state), it may be more informative to understand how the importance of model parameters changes across time. This necessitates visual tools to display this information.

\smallskip

\noindent
{\bf Sensitivity functions.} For time-dependent models defined, e.g. an ODEs, the sensitivities can be displayed on a plot showing the value of an index plotted against time \cite{Olsen2019}. 

\smallskip
 
\noindent
{\bf Phase portrait plots.} This is another way to visualise the sensitivities of time-dependent ODE models. For example, after using the Morris method, we can produce a plot showing $\mu^*$ against $\sigma$ at different time-points, drawing arrows to indicate the direction of evolution (see \cite[Figure 4]{Wainwright2014a}). Also, for the Sobol method we can plot $S_{T_i}-S_i$ against $S_i$ for different time-points, identifying the ratio between the first-order and the interaction effects between parameters. (see \cite[Figure 5]{Wainwright2014a}). Recall that all the parameters for which $S_{T_i}-S_{i}$ is greater than zero show interaction effects.

%%%%%%%%%%%%%%%%%%%%%%%%%%%%%%%%%%%%%%%%%
\section{Software for performing SA}
Software packages for performing SA are available in several programming languages. 

\medskip

\noindent
{\bf Dakota} \cite{Dakota2019}. Dakota is a software package that performs, among other applications, global sensitivity analysis using methods such as the Morris and Sobol.  The software has been applied, for instance, to conduct SA on models in immunology \cite{Salim2016}.

\medskip

\noindent
{\bf Data2Dynamics} \cite{Raue2015}.  A MATLAB package, based on a previous toolbox known as SBToolbox2, which performs parameter estimation of ODE models as well as uncertainty analysis and local sensitivity analysis. It is especially suitable for models of biochemical reaction networks. 

\medskip

\noindent
{\bf DyGloSA} \cite{Baumuratova2013}.  A MATLAB toolbox that performs global SA of models defined as differential equations, particularly focusing on detecting the critical transitions that occur in dynamical models with bifurcations.  

\medskip

\noindent
{\bf GUI-HDMR} \cite{Ziehn2009}. A MATLAB-based software that includes a set of tools to construct a metamodel and calculate variance based SA. It comes with a GUI, but can also be used as a script. The software has been applied to cyclohexane oxidation \cite{Ziehn2009a} and genetic transcription \cite{Dresch2010}. 

\medskip

\noindent
{\bf PeTTSy} \cite{Domijan2016}.  A MATLAB toolbox which implements different techniques for the perturbation theory and SA aimed at large and complex ODE models.

\medskip

\noindent
{\bf PSUADE} \cite{Gan2014}.  A C++ package for, among other application, global SA and includes techniques such as the Morris and FAST methods. It was originally developed for models in hydrology \cite{Zhan2013}.  

\medskip

\noindent
{\bf R Packages.} A number of packages for computing SA indices can be found for users of R, including {\it sensitivity} \cite{Pujol2014} and {\it ODEsensitivity} \cite{ODEsens2018}. The package {\it sensitivity} implements  global SA methods such as the Morris, FAST and Sobol methods, using different schemes for numerical computation such as the Jansen and Saltelli implementations, for instance (see Table 2) \cite{Pujol2014}. The Sobol and Morris methods also support models with multivariable outputs. {\it ODEsensitivity} allows users to conduct sensitivity analyses on ODE models using the Morris and Sobol methods.

\medskip

\noindent
{\bf SAFE} \cite{Pianosi2015b}. A MATLAB/Octave toolbox for global SA including Morris, Sobol and FAST as well as visualisation tools such as scatterplots and parallel coordinate plots. The toolbox has been applied to hydrological and climate models \cite{Pianosi2016, Almeida2017}.

\medskip

\noindent
{\bf SaLib} \cite{Herman2017}. A Python library for performing SA. It contains a number of methods including Sobol, Morris, FAST, Moment-Independent Measure, DGSM and Fractional Factorial Sensitivity Measure. 

\medskip

\noindent
{\bf SaSAT} \cite{Hoare2008}. A package for sampling and SA tools. While built using MATLAB, it can be run as a standalone program. The package contains the Sobol and regression methods for the purpose of factor ranking. 
% **Adam, could you please try installing this one?**

\medskip

\noindent
{\bf SBToolbox2}  \cite{Schmidt2006}.  A GUI-based MATLAB toolbox allowsing users to implement and simulate models, as well as to conduct bifurcation and identifiability analyses, parameter tuning and SA using local as well as global methods such as the Sobol, eFAST and the partial rank correlation coefficients. The toolbox has been used for SA of models in systems biology \cite{Williamson2012, Yi2007, Mack2014}.

\medskip

\noindent
{\bf SensSB} \cite{Rodriguez-Fernandez2010}.  A MATLAB toolbox aimed at covering main steps involved during the modelling process including sensitivity and identifiability analysis, sensitivity-based optimal design and parameter estimation. The toolbox offers three global SA methods: the Sobol and Morris methods  and DGSM \cite{Kucherenko2009}. 
%**This package is no longer maintained - it is still downloadable but contains .p files and thus require older matlab versions to run**

\medskip

\noindent
{\bf SIMLAB}  \cite{Tarantola2017}. A GUI-based MATLAB and Fortran software incorporating global and regression SA techniques, including the Morris, Sobol, FAST and eFAST methods for the former, as well as the Pearson correlation coefficient, Spearman rank coefficient and standard and partial regression coefficients for the latter. The model factors can also be specified to take common distributions such as the uniform, normal and exponential. The software has been used by researchers examining, for instance, building performance \cite{Hopfe2011}, and environmental policy \cite{Tarantola2002}.

\medskip

\noindent
{\bf SobolGSA} \cite{SobolGSA}. A GUI-based MATLAB package that implements global SA techniques for factor identifiability, including the Morris, Sobol and eFAST methods for factor ranking. The package has been used to implement SA on models investigating, for example, the management of water systems \cite{Koleva2018, Guerra2018}.

%%%%%%%%%%%%%%%%%%%%%%%%%%%%%%%%%%%%%%%%%
\section{Framework for applying SA}
SA is a part of model analysis (see Figure\,\ref{FlowChart}) and thus can be used to modify the model and  generate new hypotheses about the phenomenon being modelled. The objective is to implement the SA carefully, bearing in mind the aim of the analysis. To do this requires the selection of the appropriate SA method(s), as well as implementation and interpretation of the analysis. With this in mind, we provide here a brief guide discussing the issues to consider and pitfalls to avoid.

%---------------------------------------------------
\subsection{A step-by-step application of SA}
From a top-down point of view, the selection of appropriate SA methods can be seen as a step-by-step process. We describe each of these steps below.

\medskip
	
\noindent
{\bf Step 1: Define the purpose of the SA.}
The first step in choosing an appropriate SA method is to clearly define the purpose of the SA (e.g. factor ranking or screening).
While most approaches reviewed here can be used for factor ranking, relatively few are suitable for screening. 
 Additionally, the SA practitioner should consider the type of output measures to be investigated. For example, while it is common to use the output value at the final time \cite{Laranjeira2017, Laranjeira2018, Tanaka2011} or over time \cite{Wainwright2014a,Wu2013,Bianca2012}, features such as the ratios of multiple outputs, peak concentration time or the number of oscillations in a period may provide more insight for biological applications. 
 
\noindent
{\bf Step 2: Choose input factors, their ranges and distributions.}
When applying a concrete sensitivity method to a concrete model, the SA practitioner must select a group of parameters and their ranges (i.e. the parameter or fact space) within which the SA is to be performed. For the local SA methods this will be around a point of special interest, e.g. a steady-state of an ODE model, while for global methods a region of the parameter space, typically a neighbourhood of the nominal value. For models in physics, this value could be a measured quantity (e.g. mass, temperature). However, when the model has a biomedical focus, determining the nominal values or regions is usually non-trivial. In practice, many of the values can be determined by measuring the physical quantities by field experts through consensus or by preliminary parameter estimation (the inverse problem). Also to be considered is the distribution of the input ranges. Again, this is harder to discern for biomedical models but, if unclear, a uniform distribution is usually assumed. Care must also be taken when defining the parameter space under SA consideration as its size may influence the final ranking of the factors, which is one of the known pitfalls in applying SA. 

\smallskip

\noindent
{\bf Step 3: Select SA method(s).} 
The selection of the appropriate SA method(s) constitutes the next step. Two points are to be considered here: the model structure and computational cost of each method. At the heart of the first point is whether or not the input-output relationship of the model is linear. Whereas all SA methods can handle linear and near-linear models, only a few - all of which are global - are suitable for analysing nonlinear models, or those with interactions between factors. The computational cost of using each method is the other point to consider. In general, when applying SA, the number of model evaluations increases with the number of model factors. The exact tread-off depends on the approach and the model itself \cite{Pianosi2016}. Although variance-based methods (e.g. Sobol) are the preferred method for many SA applications, their use may be limited by the model's complexity. This relationship between computational cost and quantitative detail of the SA method applies to all other methods too. For instance, local methods are computationally cheaper than global ones, but provide less information away from the operating point.  

\smallskip

Table \ref{MethodsSummary} provides a summary of the SA methods examined in this review and their properties. It is intended to be used as a guide to the selection of appropriate SA methods.

\smallskip

If possible, it may also be useful to compare the results of more than one SA method and, if appropriate and computationally feasible, to use variance-based approaches. Some examples in the literature of combining SA methods are available \cite{Jarrett2017,Wainwright2014a,Link2018}. However, as is common in biomedical sciences, models in this field tend to involve large number of equations and input parameters and therefore the application of SA methods such as the Sobol to the full model is computationally prohibitive. In this case, one option would be to start with computationally efficient screening methods such as the Morris and subsequently apply more intensive variance-based methods to a selected set of parameters. However, if this is not possible, settling with even one simple sensitivity method may still be revealing for the problem at hand \cite{Laranjeira2017, Laranjeira2018}.

\noindent
{\bf Step 4: Visualisation and interpretation}.
The message to be conveyed through the application of SA can be emphasised by visualisation of the results. There is a large range of graphical tools available to the SA practitioner; these are discussed in Section \ref{GraphicalSection}. Finally, the results of the SA are interpreted according to the specificity of the biomedical and biological application. Typical points to consider may include model selection, model reduction and practical identifiability.

%-----------------------------------------------------------------
\subsection{Pitfalls in the application of SA}  \label{OATLimitationsSection}
Although SA is an important part of model analysis, the inferences we derive may be misleading or even incorrect if the SA is perfunctory. Some common errors are outlined in this section. 

\smallskip

\noindent
{\bf Indiscriminate use of OAT methods.} Although popular, OAT methods explore only a small fraction of the possible set of parameter values (the parameter space), especially when the number of model parameters is high \cite{Saltelli2019b}. For example, in the case of a 12 parameter model, an OAT method will explore less than one-thousandth of the parameter space \cite{Saltelli2010a}. Hence, the simplicity of these methods comes at the cost of a more thorough understanding of the parameter space. 

 Also, because only a small proportion of inputs has an influential role on the output \cite{Saltelli2008}, no matter how large the model, OAT approaches are also considered to be inefficient when compared with their MAT counterparts. Hence, the use of OAT approaches results in most model evaluations adding little further information. Also, by varying only one input, it is difficult to ascertain the effect  interactions have on the output uncertainty \cite{Saltelli2008}. While some OAT techniques, such as the Morris method, do tell us whether or not input interactions exist, this information is only semi-quantitative. We are given only an index showing the importance of the input's interactions but no information about, for instance, with which particular inputs these interactions are formed \cite{Saltelli2008}. 

\smallskip

\noindent
{\bf Insufficient sampling size.} The number of samples, $N$, in sampling-based approaches may affect the convergence of the sensitivity indicies \cite{Wang2013, Tarantola2012}. The larger the sample size, the more accurate the results will be but this involves a trade-off with the computational cost required to run the method on these extra samples. The error inherent in the Sobol method, for example, is proportional to $1/ \sqrt{N}$ \cite{Homma1996}. Moreover, if the sample size is too small, we may obtain inaccurate results. For instance, the occurrence of negative Sobol indices (which should not arise in an ideal setting) suggest large approximation errors during the estimation of the indices and these can only be reduced by increasing the sample size. 

\smallskip

\noindent
{\bf Use of correlation methods for inappropriate input-output relationships.} Another pitfall involves the use of correlation methods for models with inappropriate input-output relationships. As noted in Section \ref{corregress} and in \cite{Schober2018}, the correlation measures $\CC_i$, $\SRC_i$ and $\PCC_i$ require a linear relationship between the model input and output, while monotonicity is neccesary for the use of $\SRCC_i$ and $\PRCC_i$. If the appropriate conditions are not satisfied, then the results of the correlation analysis will be unreliable. 

\smallskip

\noindent
{\bf Non-biological parameter ranges.} As discussed earlier, global SA techniques (e.g. Morris and Sobol) require the parameter ranges to be defined. Changing the ranges may affect significantly the sensitivity indices and potentially lead to insensitive parameters becoming sensitive, or vice versa.
Another common occurrence in ODE models in the biomedical field is the qualitative change in model dynamics when crossing certain regions of the parameter space (i.e. bifurcations). This can include the formation of new steady states (as seen, for instance, in infectious disease models, where each equilibrium represents a healthy or weakened immune response to the disease \cite{Stan2008}), changes in a model's stability \cite{Swat2004} and creation of periodic orbits (e.g. calcium ions in a human cell \cite{Domijan2006}). SA methods do not differentiate between these different regions of the parameter space and so will return some average of these qualitatively different dynamics \cite{VanVoorn2017}.
For biomedical models, it is important that the ranges are biologically feasible so that the sensitivity indices are not biased by implausible model realisations  \cite{Shin2013}.

\noindent
{\bf Using local methods for nonlinear models.}
A frequently-encountered pitfall arises when we use local methods for nonlinear models. To illustrate this, we apply derivative-based local sensitivities to model \eqref{ToyModel}, but this time at a different point, namely where all factors take a value of $0.9$ (instead of $0.1$).  We show that when the model is nonlinear, then the rankings produced by these methods may depend on the point in the parameter space. Table \ref{TableLocalPoint9} displays the results of the derivative-based local SA when all factors are set to $0.9$. We can observe the swapping of the order of importance of factors $X_1$ and $X_4$, compared with Table \ref{TableLocal}, where all factors are set to $0.1$. This reinforces our statement that, while a local SA provides a simple glimpse into the importance of a model's factors, the results are meaningful only at the point at which the sensitivities are calculated.

\begin{table}[t!]
\centering
\begin{tabu}{lccc}
\tabucline[0.7pt] {-}
Factor & 	$S_i^{\rm abs}$									&	$S_i^{\rm rel}$						      &$S_i^{\rm var}$	\\
\hline
$X_1$  & 1.8                                                                                         	& 0.64                                                                        & 0.130                                                                                     \\
$X_2$  & 0.9                                                                                          	& 0.32                                                                        & 0.064                                                                                        \\
$X_3$  & 0.9                                                                                          	& 0.32                                                                        & 0.064                                                                                        \\
$X_4$  & 1.0                                                                                              	& 0.35                                                                        & 0.071                                                                                        
\\
\tabucline[0.7pt] {-}
\end{tabu}
\caption{This table shows the results of a local derivative-based sensitivity applied to model \eqref{ToyModel}. We present, here, the absolute, relative and variance-based sensitivity indices, measured when all factors have values of 0.9. For the relative sensitivity measures, we choose $X_i^0=0.9$ and $Y_i^0=2.52$. For the variance-based sensitivity, we calculate that $\sigma_{X_i}^2=1/12$ and $\sigma_Y^2=14/12$.}
\label{TableLocalPoint9}
\end{table}

%-----------------------------------------------------------------
\subsection{Robustness of SA estimates}
All SA methods are subject to uncertainty. Robustness analysis assesses the dependence of sensitivity indices with respect to the quantities chosen during the computation, such as the distribution of the model parameters (factors), their correlation structure or the choice of a specific sample and its size. 

\begin{figure}[t!]
\centering
\includegraphics[width=\linewidth]{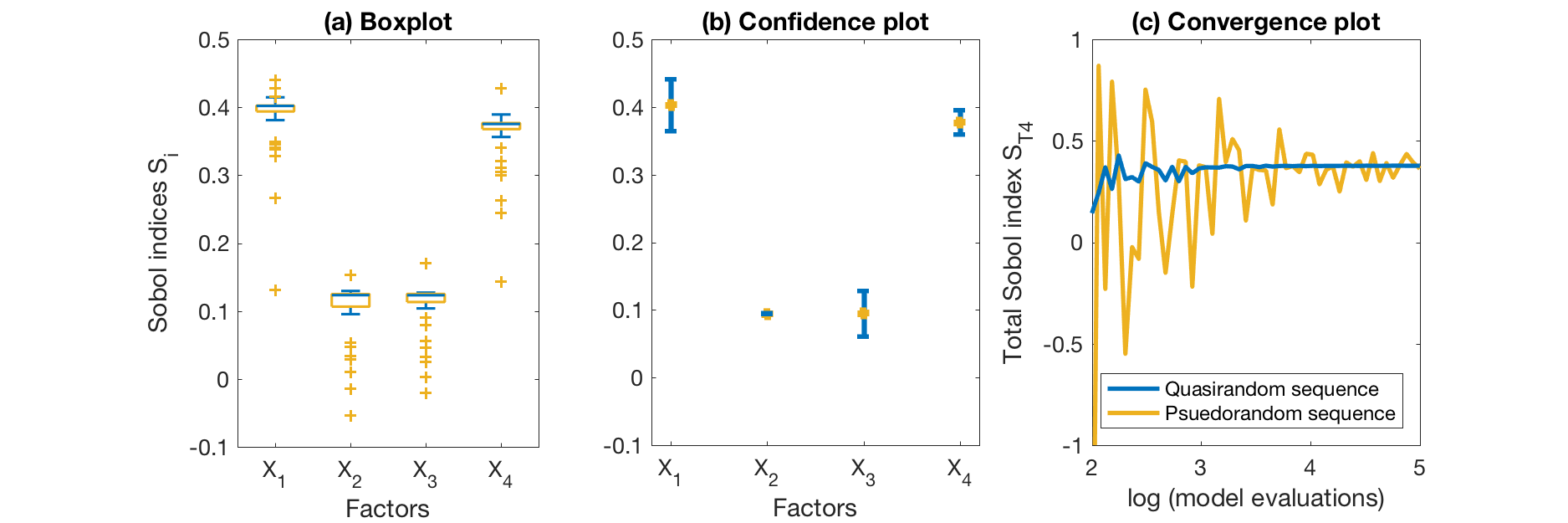}
\caption{Graphical methods for robustness of SA indices.}\label{fig:all_rel}
\end{figure}

The robustness of Sobol indices to changes in the distributional uncertainty has been a focal point of several recent studies \cite{Paleari2016, Hart2018, Hart2018a}. This may have important consequences since it has been shown that changes in the marginal distributions may change the ordering of the Sobol indices in \cite{Cousin2019}.
In sampling based methods, the most basic question is to understand how many samples do we need for the given SA index to achieve the convergence \cite{Iman1982, Iman1990}. A popular qualitative assessment is to plot sensitivity against the number of evaluations \cite{Nossent2011},  as shown in Figure\,\ref{fig:all_rel}. Convergence and uncertainty analysis based on the Central Limit Theorem has been discussed in \cite{Yang2011}. However, this is only computationally feasible for relatively simple models. For complex models, methods based on bootstrap techniques may be more suitable \cite{Efron1994, Archer1997}. 

The popular visual methods for robustness include the boxplot, confidence interval and convergence plot. The {\it boxplot}, also known as a {\it box-and-whisker plot}, displays the medium, upper and lower quartiles of the model's sensitivity indices across several bootstrap resamples. Figure\,\ref{fig:all_rel}a shows a boxplot illustrating these quartiles for the Sobol sensitivity indices of the example model \eqref{ToyModel}.  {\it Confidence interval plots} can be used to illustrate the uncertainty of the sensitivity measures using, e.g. bootstrap resamples and preselected confidence intervals \cite{Hoare2008}. In \cite{Archer1997} the authors introduced a method to calculate the symmetric 95\% bootstrapped confidence intervals of the Sobol indices. Figure\,\ref{fig:all_rel}b shows a confidence interval plot created using this method to illustrate the relevant confidence intervals for the Sobol indices of the example model \eqref{ToyModel}. Figure\,\ref{fig:all_rel}c shows a {\it convergence plot} illustrating the number of model evaluations required for the Sobol indices calculated for the example model \eqref{ToyModel} to converge to their true values. Here, we compare the convergence of the Sobol indices computed using two sampling sequences: pseudorandom and quasi-random (specifically, a Sobol sequence). Upon comparison, we can see that using quasi-random sequences enables faster convergence. Notice, also, that the Sobol indices tend incorrectly to give negative values if insufficient model evaluations are computed. 
 
%-----------------------------------------------------------------
\subsection{Correlated input factors}
It is common in practical applications that the input factors will be, at least to some extent, correlated. A number of challenges arise in performing sensitivity analysis for correlated inputs \cite{Jacques2006, Xu2008, DaVeiga2009, BI2012, Gromping2007, Strobl2008}. The first-order and higher-order Sobol indices  (see Section \ref{sobol}), are related through the variance decomposition \cite{Sobol1993}, which holds only if the input factors are independent \cite{Oakley2004}. The case of dependent factors has been considered separately in \cite{Saltelli2002c}. Unexpected results may arise if factors are correlated. One may find that the first-order indices are higher than the total-order indices, which may depend on the level of correlation; or total-order indices tending to zero as the correlations approach one \cite{Kucherenko2012}. In the context of regression methods (Section \ref{corregress}), collinearity may produce large variances of some estimated regression coefficients \cite{Wei2013}, which obscures our ability to interpret the results of the analysis.

%%%%%%%%%%%%%%%%%%%%%
\section{Application of SA in biomedical sciences}

%---------------------------------------------------
\subsection{Algebraic model: cancer prediction}

In this section, we will illustrate two commonly-used global SA techniques the Morris and Sobol methods by applying them to a model of colorectal cancer \cite{Calabrese2010}. The colon, or large intestine, forms the final stage of the digestive tract. The lining of this colon is frequently replaced by cells in a gland known as the colonic crypt but occasionally mistakes are made during this process. If such a mistake leads to the new, mutant cells growing and dividing abnormally, the growth may become cancerous.  

The authors of \cite{Calabrese2010} present a simple model by assuming that only five factors cause cancer. The model is as follows
\begin{equation}\label{Cancer:algmodel}
p = 1-\big(1-(1-(1-u)^d)^k\big)^{nm},
\end{equation}
where $p$ is the probability that an individual is affected by cancer, $u$ the mutation rate of the cells per division, $d$ the number of divisions, $k$ the number of rate-limiting mutations required for cancer to occur, $n$ the number of stem cells per crypt and $m$ the number of crypts in the colon. Table \ref{CancerPrediction} shows the factors and their maximum and minimum values. Here, we assume that the individual is 70 years old, giving the appropriate values of the factor $d$.

In \cite{Calabrese2010}, two sets of parameter values are given. The first corresponds to the theory that mutations due to cancer tend to occur in specific sets of susceptible genes. Another school of thought suggests that, in contrast, functional or regulatory pathways are more susceptible to cancer \cite{Sjoblom2006, Wood2007, Jones2008, Parsons2008}; the second set of parameter values reflects this. 
We will assume that these two parameter sets correspond to, respectively, the minimum and maximum values of the parameter range, and that all the resultant parameter ranges follow a uniform distribution. 

\begin{table}
\centering
\begin{tabu}{llll}
\tabucline[0.7pt] {-}
Factor 	& Physiological Representation                 		 	& Min Value   		& Max Value                     \\
\hline
$u$   	& Mutation rate                                 				&  $10^{-6}$         		& $ 3 \times 10^{-6} $ 		\\
$d$   	& Divisions per cell during lifetime				& $6 \times 10^{3} $ 		& $2 \times 10^4$                      \\
$k$   	& Rate-limiting mutations                          			& 5               			& 6                                              \\
$n$   	& Number of stem cells per crypt                		& 8               			& 40                                            \\
$m$   	& Number of crypts in colon                     			& $10^7$          			& $2 \times 10^{7}$                     \\
\tabucline[0.7pt] {-}   
\end{tabu}
\caption {The factors involved in the cancer prediction model. The minimum values given here correspond to the case with specific gene targets, and the maximum values to the pathway gene targets - values for both cases are taken from \cite{Calabrese2010}.}

\label{CancerPrediction}
\end{table}

\begin{figure}
\centering 
\includegraphics[width=140mm]{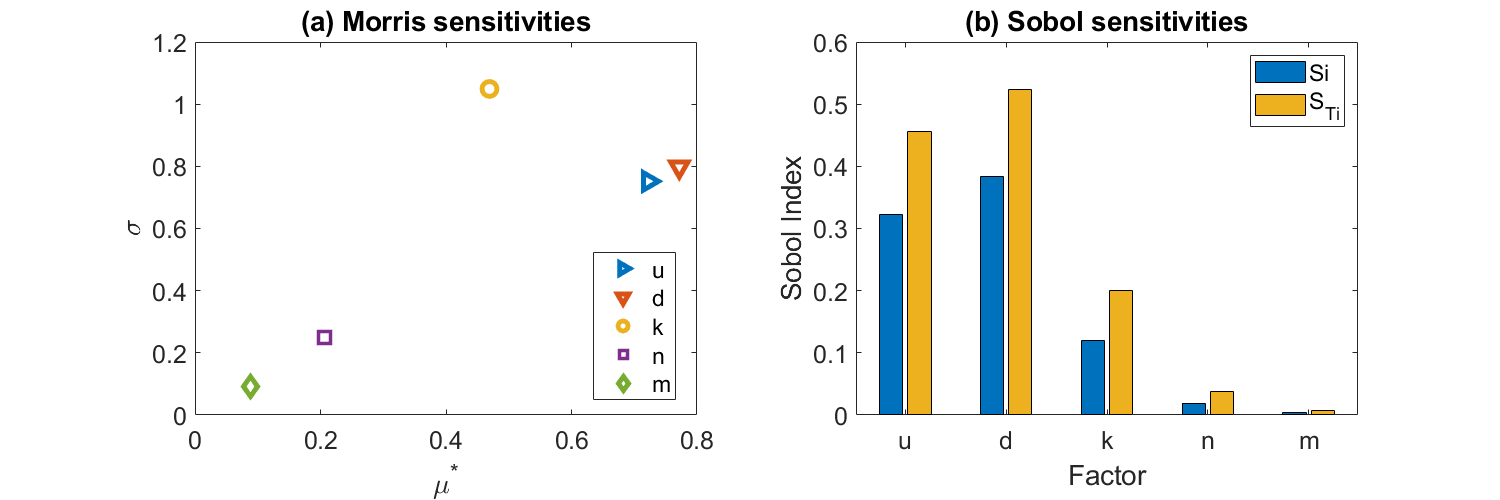}
\caption{(a) The $\mu^*$-$\sigma$ plot showing the influence of the 5 model factors on the output $p$ using the Morris method. (b) Results after running the Sobol method on the cancer prediction model \eqref{Cancer:algmodel}. Displayed are the first-order $S_i$ and total-order $S_{T_i }$ indices for each of the five model factors.}
\label{fig:CancerPredictionFigs}
\end{figure}

We investigate the impact each of the five factors has on the variation in the output by applying the Morris and Sobol methods to the model. Figure \ref{fig:CancerPredictionFigs} shows the results of the respective methods.

Both SA methods show that factors $d$ and $u$, which represent the number of cell divisions and mutation rate respectively, have the greatest effect on the output variance and are greatly involved in interactions with other factors; $k$ has moderate effect on the output but is still heavily involved in interactions. Factors $m$ and $n$, however, are deemed to have low output influence and their interactive effects are small, compared with the other factors. The intuitive mathematical explanation for this is that the factors $u$ and $d$ sit `deep' within model equation \eqref{Cancer:algmodel} and so any variations in their values will be compounded by the exponents (namely, $k$ and $n\times m$) encountered later in the equation.

From a physiological viewpoint, these results seem sensible. Bowel cancer is caused by multiple genetic mutations in stem cells \cite{Lee2010} and the two factors that most directly correspond to the number of mutations are $d$ and $u$. An important study by Tomasetti and Vogelstein \cite{Tomasetti2015} showed that the number of cell divisions over a person's lifetime was the most influential factor for explaining the difference in cancer rates between different body tissues. Hence, the authors concluded that this was the major contributing factor in cancer development. Our SA of such a simple algebraic model reaches the same conclusion; a reflection of its power and importance. 

%---------------------------------------------------
\subsection{ODE model: cell differentiation in the colon}
\label{ODECellDiffModel}
This section involves the application of the Morris and Sobol methods to an ODE model describing the differentiation of stem cells in the colon, introduced in \cite{Johnston2007a}. As cell division occurs frequently in the bowels, the chances of carcinogenesis occurring there are abnormally high \cite{Brittan2004}. The model of Johnston et al., \cite{Johnston2007a}, attempts to understand how cell differentiation and homeostasis occurs in both healthy bowels and cases where carcinogenesis occurs. 

In this model, three types of cells reside in the colon: stem cells, semi-differentiated and fully differentiated cells. Stem cells, found at the bottom of colonic crypts \cite{Preston2003}, have the ability to differentiate into specialised cells, such as Goblet cells and colonocytes for purposes such as tissue regeneration \cite{Brittan2004}. Once these stem cells have been given the signal to differentiate, they travel up along the crypt. During this phase, they are known as semi-differentiated or transit-amplifying cells. Having completed the migration up the crypt, the cells become fully differentiated into their specialisations. During the first two stages, cells are lost due to cell death, regenerated through proliferation or undergo differentiation to the next stage. Fully differentiated cells cannot proliferate (their generation comes only from differentiation of transit cells) but are still lost from their death and/or removal. In healthy individuals, all cell populations are kept in homeostasis; during carcinogenesis, however, cell differentiation is left unchecked. 

The model of Johnston et al. \cite{Johnston2007a} portrays these cell dynamics. The authors propose the following feedback model to enable the different cell populations to remain in homeostasis. The model with saturating feedback has the form:
\begin{align}
 \label{eqn1Stem}
\ \frac{dN_0}{dt} &= (\alpha_3-\alpha_1-\alpha_2)N_0-\frac{k_0N^2_0}{1+m_0N_0}\\ 
 \label{eqn2Stem}
\ \frac{dN_1}{dt} &= (\beta_3-\beta_1-\beta_2)N_1+N\alpha_2N_0-\frac{k_1N^2_1}{1+m_1N_1}+\frac{k_0N^2_0}{1+m_0N_0}\\
 \label{eqn3Stem}
\ \frac{dN_2}{dt} &= -\gamma N_2 + \beta_2N_1+\frac{k_1N_1}{1+m_1N^2_1},
\end{align}
where the state variables $N_0$, $N_1$ and $N_2$ represent the populations of stem cells, semi-differentiated and fully differentiated cells, respectively (see Table \ref{StemCell}). The initial conditions are chosen to be $N_0=1$, $N_1=100$ and $N_2=100$. These values are a reflection of the cell populations seen in the colon, where there are far fewer stem cells than their differentiated counterparts.

% \usepackage{multirow}
\begin{table}[]
\begin{center}
\begin{tabu}{lll}
\tabucline[0.7pt] {-}
Factor     		& Physiological Representation                     		& Suggested Value \\
\hline
$\alpha_1$ 	& Rate of Stem Cell Death                          		& $0.1$           	\\
$\alpha_2$ 	& Rate of Stem Cell Differentiation                		& $0.3$           	\\
$\alpha_3$ 	& Rate of Stem Cell Proliferation                  		& $0.69$          	\\
$\beta_1$  	& Rate of Semi-Differentiated Cell Death           		& $0.1$           	\\
$\beta_2$  	& Rate of Semi-Differentiated Cell Differentiation 	& $0.3$           	\\
$\beta_3$  	& Rate of Semi-Differentiated Cell Proliferation   	& $0.397$         \\
$\gamma$   	& Rate of Removal of Fully Differentiated Cells    	& $0.139$         \\
$k_0$      		& Feedback Constant              					& $0.1$           	\\
$m_0$     		& Feedback Constant                                                  & $0.1$           	\\
$k_1$      		& Feedback Constant                                                  & $0.0003$       \\
$m_1$      	& Feedback Constant                                                 & $0.0004$       \\
\tabucline[0.7pt] {-}
\end{tabu}
\caption{This table summarises the parameters of the cell differentiation model. Values given here are empirical and reproduced from \cite{Johnston2007}}.
\label{StemCell}
\end{center}
\end{table}

\begin{figure}[h!]
\centering
\includegraphics[width=0.45\linewidth]{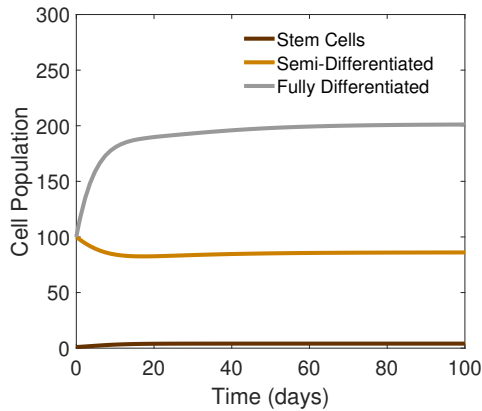}
\caption{Time-course simulations of the cell differentiation model  \eqref{eqn1Stem}-\eqref{eqn3Stem}.}\label{fig:TimeCourse}
\end{figure}

Figure \ref{fig:TimeCourse} shows the time-course plot acquired after running the model from these initial conditions. The model eventually reaches the steady-state $N_0^*=4$, $N_1^*=86$, $N_2^*=201$. This is the only equilibrium point for this particular set of parameter values \cite{Johnston2007a}.

We now conduct SA on both these models using the Morris and Sobol methods. We aim to determine which parameters and, therefore, which cell processes may be of importance to the model and for cell homeostasis. The results presented in the next two sections refer only to sensitivities with respect to the population of stem cells. We will leave a discussion on how the results change when the output of interest switches to $N_1$ and $N_2$ in the appendix. 

Finally, just as with the cancer prediction model, we note that the range of parameter values is influential to the SA. As only one nominal value for every parameter is given in \cite{Johnston2007a}, we will define each parameter's range to be a uniform distribution spanning $\pm 10\%$ of its given value. Different parameter ranges, however, will alter the analysis and so we emphasise the importance of determining these ranges accurately.

%---------------------------------------------------
\subsubsection*{Cell Differentiation Model: Morris method}
As we have seen, the Morris method outputs the mean of the absolute values of the elementary effects associated with each parameter ($\mu^*$), as well as the standard deviations of these elementary effects ($\sigma$). Applying the Morris Method to ODE models allows us to track $\mu^*$ and $\sigma$ for each parameter as time (or whatever the independent variable happens to be) progresses. For ODEs with $n$ state variables, we obtain $n$ sets of $\mu^*$ and $\sigma$ curves. Hence, both cell differentiation models will have three sets of  $\mu^*$ and $\sigma$ values, each corresponding to their influence on $N_0$, $N_1$ and $N_2$.

\begin{figure}
\centering 
\includegraphics[width=0.9\linewidth]{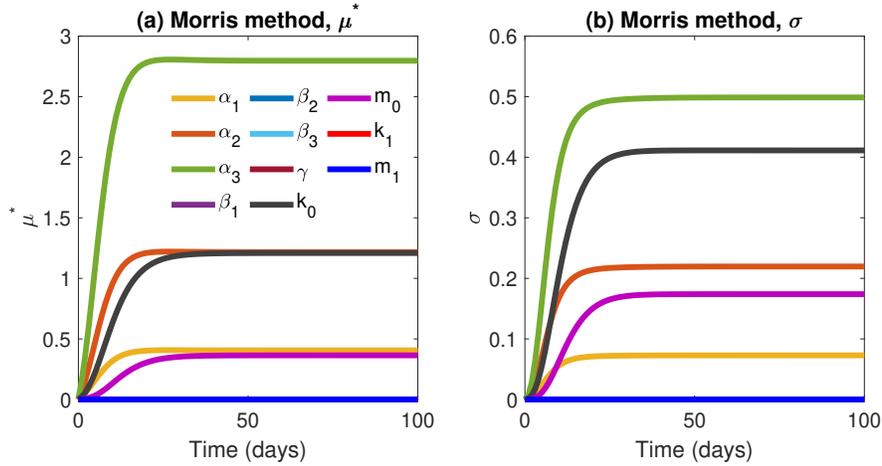}
\caption{The average of the absolute values of the elementary effects, $\mu^*$, (a), and standard deviation of the values of the elementary effects $\sigma$, (b), associated with each of the parameters in the model (equations \ref{eqn1Stem}-\ref{eqn3Stem}) using the Morris method as functions of time.}
\label{fig:MorrisODEFigs}
\end{figure}

The $\mu^*$ and $\sigma$ values, as functions of time, of the cell differentiation model, with respect to state variable $N_0$, are shown in Figures \ref{fig:MorrisODEFigs}a and \ref{fig:MorrisODEFigs}b. The Morris method predicts that only five of the eleven model parameters have any non-negligible effect on the output variation of $N_0$ (positive $\mu^*$) and interact with other parameters (positive $\sigma$). 
Inspecting the $\mu^*$ values in Figure \ref{fig:MorrisODEFigs}a, the parameter that most affects the output variance is $\alpha_3$, which represents stem cell proliferation. Two more parameters, $\alpha_2$ and $k_0$, representing stem cell differentiation in the presence of feedback, also have moderate effects on the output, while the only other parameters with non-negligible influence on $N_0$ are $\alpha_1$ and $m_0$. All other parameters are deemed not to influence $N_0$. Figure \ref{fig:MorrisODEFigs}b shows the $\sigma$ values of the eleven parameters. Again, $\alpha_3$ and $k_0$ are the parameters that participate most in interactions with parameters. $\alpha_2$, $m_0$ and $\alpha_1$, too, have non-negligible $\sigma$ values but the remaining parameters have few interactions with one another. The values of $\mu^*$ and $\sigma$ for all parameters reaches peaks at around 10-20 days.

We can understand why only five parameters affect the output $N_0$ by simply inspecting equation \eqref{eqn1Stem} and noticing that it can be decoupled from the rest of the model. Hence, only the parameters found in this equation can affect the output; since $N_0 (t)$ changes whenever the parameters change, the mean of the elementary effects absolute values ($\mu^*$) must be positive for each parameter. 
The standard deviation of the elementary effects caused by each of the 5 parameters must also be greater than zero. As noted previously, this implies that the parameters must interact with each other or participate in nonlinear effects and yet this is not immediately clear from equation \eqref{eqn1Stem}: from this perspective, it appears that $\alpha_1$, $\alpha_2$ and $\alpha_3$ are linear terms that do not interact with one another. However, what we must remember is that we are not interested in parameter interactions in the differential equation {\it per se}, but, rather, their nonlinearities or interactions with each other in the integral of equation \eqref{eqn1Stem}, i.e. in the solution $N_0 (t)$. One way to think about this is to take only the linear part of the equation:
\begin{equation}
\begin{aligned}
\frac{dN_0}{dt}=(\alpha_3 - \alpha_1 - \alpha_2 ) N_0,
\end{aligned}
\label{eqn:solutioneqn}
\end{equation}
which has the solution  $N_0 (t)= A \exp (\alpha_3 - \alpha_1 - \alpha_2) t$, for some constant $A$. Hence, incrementing any of the three parameters here, even though they appear in a linear differential equation, has a nonlinear effect on the solution due to the exponential function.

%---------------------------------------------------
\subsubsection*{Cell Differentiation Model: Sobol method}

\begin{figure}
\centering 
\includegraphics[width=0.9\linewidth]{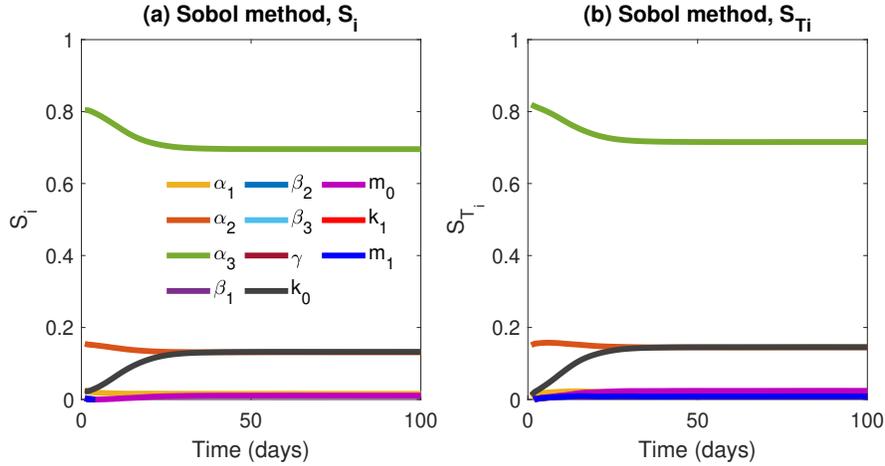}
\caption{The first-order Sobol indices, (a), and total-order Sobol indices, (b), associated with each of the parameters in the model (equations \ref{eqn1Stem}-\ref{eqn3Stem}) as functions of time}
\label{fig:SobolODEFigs}
\end{figure}

Here, the Sobol method is applied to the ODE model to obtain Sobol's  first-order indices $S_i$ as well as the total-order sensitivity indices $S_{T_i}$ as a function of time. Just as in the case of the Morris, applying the Sobol method to ODEs with $n$ state variables returns $n$ sets of $S_i$ and a further $n$ sets of $S_{T_i}$ values. Hence, both cell differentiation models will have three sets of $S_i$ and $S_{T_i}$, each corresponding to their influence on $N_0$, $N_1$ and $N_2$. Here, we concentrate on the influence of the parameters on state variable $N_0$. Figures \ref{fig:SobolODEFigs}a and \ref{fig:SobolODEFigs}b show the results, with respect to this variable, after applying the Sobol method to the cell differentiation model.

According to the total-order sensitivity indices, parameter $\alpha_3$ is the most influential. $\alpha_2$ and $k_0$ also contribute to the output variance of the saturating feedback model, but to a much lesser extent, while $\alpha_1$ and $m_0$ have a very small, though positive. We notice that the rankings produced by the total-order sensitivity indices matches those predicted by the Morris methods $\mu^*$, as should be the case \cite{Campolongo, Campolongo2011}. What the total-order Sobol indices do add, however, is that they quantify exactly each parameters contribution of the output variance; the $\mu^*$ of the Morris method, being a semi-quantitative method, can only display a ranking of parameter importance, though it is particularly effective at identifying unidentifiable parameters.   

We can gain even more information by using both the Morris and Sobol methods in conjunction. While the $\sigma$ values from the Morris method provide an indication of both the nonlinearities caused by a parameter as well as the interactions it is involved in, the difference between its total and first-order Sobol indices quantifies only the effects of the parameter's interactions \cite{Wainwright2014a}. We can, therefore, separate a parameter's nonlinear and interaction effects. Inspecting this example, we observe that the $\sigma$ values of many parameters are non-zero (Figure \ref{fig:MorrisODEFigs}) and so they must produce nonlinear and/or interaction effects. Now, we combine this knowledge with our observations from Figure \ref{fig:SobolODEFigs}, which shows little difference between the first and total-order Sobol indices of each of the 11 parameters. Hence, we conclude that the vast majority of the $\sigma$ values of these parameters can be attributed to their nonlinear effects, rather than their interactions with other parameters. This can be explained, again, by inspecting \eqref{eqn:solutioneqn} and noting that the relevant parameters affect the solution through nonlinearities rather than interactions.

%---------------------------------------------------
\subsubsection*{Cell Differentiation Model: physiological interpretation}
Having explained the results of the sensitivity analyses from a mathematical perspective, we now discuss their physiological implications. We consider the stem cell population, $N_0$. As discussed previously, only five parameters have any effect on this population, by far the most influential being the rate of proliferation, $\alpha_3$. Smaller influences come from cell differentiation, $\alpha_2$ and the feedback term $k_0$. This signifies that changing the proliferation rate alters the stem cell population more than changes in the rates of death $\alpha_1$ and differentiation $\alpha_2$ combined an interesting observation. 

Also of interest is the fact that stem cell proliferation continues to have some, albeit much smaller, effect on the variance of the semi and fully differentiated cell populations. The rate at which the semi-differentiated cells proliferate $\beta_3$ is the most influential parameter not only for variations in the semi-differentiated cells but even for the fully-differentiated population as well. This is counter-intuitive. One might initially suppose, from the physiology, that any variance in the numbers of fully-differentiated cells would come predominantly from their removal ($\gamma$) or the transformation of semi-differentiated cells ($\beta_2$). Yet, the SA shows otherwise.  The variance of the proliferation rate of transit cells and, thus, the population of these cells, affect the variance of the fully differentiated cells even more than those two parameters. 

Such an example illustrates, again the power of conducting a SA. To conclude, we mention yet another benefit. The results of a SA on ODE models give us a good indicator of which parameters we can make suitable estimates, as well as when to make these. If we consider Figure \ref{fig:MorrisODEFigs}, for instance, we can see that we can estimate the value of $\alpha_3$ by measuring $N_0$, as the latters sensitivity to the former is high. To a lesser extent, this is also true of the parameters $\alpha_2$, $k_0$, $\alpha_1$ and $m_0$, though it would be harder to estimate these values as $N_0$ is less sensitive to them. In addition, the figure informs us to take these measurements when the model has reached steady-state, as the sensitivities are highest then. To estimate all other model parameter values, we must take measurements from $N_1$ and $N_2$, as they are unidentifiable from $N_0$.

%%%%%%%%%%%%%%%%%%%%%%%%%%%%%%%%%%%%%%%%%
\section{Conclusions}
This paper reviews SA methods, current software for performing SA and provides a framework for the analysis with the focus on biomedical sciences and biology. We also perform global sensitivity analysis on two models (an algebraic and also an ODE model) in the area of cancer biology. The intended audience for this paper is the mathematical modeller working on biomedical applications. The paper emphasises the particular importance of SA for models in biology and medicine, stemming from the stochastic nature of biological processes, uncertainty in acquired data and the need to fit models to these data. 

Although straightforward, local SA techniques consider the sensitivities at only a small region of parameter space and the conclusions derived from such an analysis are, therefore, meaningful only for linear models. Since nonlinear models are prevalent in systems biology, the use of global SA methods is usually more appropriate. The purpose of the SA must also be taken into consideration when choosing an appropriate method. Two common aims are factor ranking and screening. Having performed the analysis, we must consider both how best to visualise the results and interpret them. These include robustness of the SA estimates as well as whether or not the results have been influenced by bifurcations, such as changes in model stability and the appearance of periodic orbits; such phenomena frequently arise in the biomedical field and so must be accounted for. Finally, it is necessary to provide a biological interpretation of these SA results. 
  
It is widely acknowledged that SA is an essential part of the mathematical modelling process \cite{Saltelli2008,Saltelli2000,Rabitz1989}. Even so, there appears to be a lack of `good practice' involved in many such analyses \cite{Saltelli2019b}. This guide is intended to inform the reader of the importance of SA and provide a proper framework for its application.

%%%%%%%%%%%%%%%%%%%%%%%%%%%%%%%%%%%%%%%%%
\section{Acknowledgements}
We thank Dr Adam Szmul (University College London) for his insightful comments.

\smallskip

\noindent
In memory of Jane Yu.

%\bibliographystyle{abbrv}
%\bibliography{my_bib22}

\end{document}